\documentclass[structabstract]{aa}  
\usepackage{graphicx}
\usepackage{textcomp}
\usepackage{graphics}
\usepackage[usenames,table]{xcolor}
\usepackage{txfonts}
\usepackage{url}
\usepackage[]{natbib}
\bibpunct{(}{)}{;}{a}{}{,}
\usepackage{color,hyperref}
\definecolor{darkblue}{rgb}{0.0,0.0,1.0}
\hypersetup{colorlinks,breaklinks,
           linkcolor=darkblue,urlcolor=darkblue,
           anchorcolor=darkblue,citecolor=darkblue}
\DeclareTextSymbol{\degre}{OT1}{23}

\newcommand{\corot}{\emph{CoRoT}}

\newcommand{\flames}{\emph{FLAMES}}

\newcommand{\giraffe}{\emph{GIRAFFE}}

\newcommand{\matisse}{\emph{MATISSE}}

\newcommand{\sn}{SNR}
\newcommand{\kms}{km\,s$^{-1}$}

\newcommand{\exodat}{\emph{Exo-Dat}}

\newcommand{\vrad}{$V_{\rm rad}$}

\newcommand{\teff}{T$_{\rm eff}$}
\newcommand{\logg}{log{\it~g}}
\newcommand{\met}{[M/H]}
\newcommand{\alf}{[$\alpha$/Fe]}

\newcommand{\jmk}{(J-K_{s})}
\newcommand{\vmj}{(V-J)}

\graphicspath{{/EPSF/}{figures/}{./}}

\begin{document}

\title{Characterisation of the Galactic {  thin disc} with \corot~targets
\thanks{Tables 2 and 3 are only available in electronic form
at the CDS via anonymous ftp to cdsarc.u-strasbg.fr (130.79.128.5)
or via http://cdsweb.u-strasbg.fr/cgi-bin/qcat?J/A+A/}
}


\authorrunning{J.-C. Gazzano et al.}

  \author{J.-C.~Gazzano\inst{1,2}
         \and G. Kordopatis\inst{2,3}
	\and M.~Deleuil \inst{1}
         \and P.~de~Laverny\inst{2}
         \and A.~Recio-Blanco\inst{2}
	\and V.~Hill\inst{2}
				}
\offprints{Patrick de Laverny, \email laverny@oca.eu }
  \institute{Laboratoire d'Astrophysique de Marseille (UMR 6110), OAMP, Universit\'e Aix-Marseille \& CNRS, 38 rue Fr\'ed\'eric Joliot Curie, 13388 Marseille cedex 13, France
               \and
        Laboratoire Lagrange (UMR 7293), Universit\'e Nice Sophia Antipolis, CNRS, Observatoire de la C\^ote d'Azur, BP 4229, 06304 Nice, France
               \and
        Institute of Astronomy, University of Cambridge, Madingley Road, Cambridge CB3 0HA, UK
 }\date{Received; accepted}

 \abstract
 {}
{We use kinematical and chemical properties of 754 \corot~stars  to characterise the stellar populations  of the Milky Way disc in three beams close the Galactic plane.}
{From the atmospheric parameters derived in Gazzano et al. (2010) with the \matisse~algorithm, we derived stellar distances using isochrones. Combining these data with proper motions, we provide the complete kinematical description of stars in three \corot~fields. Finally, we used kinematical criteria to identify the Galactic populations in our sample and study their characteristics, particularly their chemistry.}
{Comparing our kinematics with the Besan\c con Galactic model, we show that, within 3$\sigma$, simulated and observed kinematical distributions are in good agreement. We study the characteristics of the thin disc, finding a correlation that is significant at a value of 2-$\sigma$
between the $V$-velocity component and the metallicity for two different radial distance bins (8-9~kpc and 9-10~kpc; but not for the most inner bin 7-8~kpc, probably because of the uncertainties in the abundances)
 which could be interpreted as radial migration evidence. We also measured a radial metallicity gradient value of    $-0.097\pm0.015$~dex~kpc$^{-1}$ with giant stars,  and    $-0.053\pm0.015$~dex~kpc$^{-1}$with dwarfs. 
Finally, we identified metal-rich stars with peculiar high \alf~values {in the directions pointing to the inner part of the Galaxy}.
Applying the same methodology to the planet-hosting stars detected by \corot~shows that they mainly belong to the thin disc population with normal chemical and kinematical properties.}
{}


\maketitle
%
\section{Introduction}
Understanding the history of the Milky Way requires a thorough study of the Galactic populations. The structure and chemistry of the Galactic thin and thick discs have already been extensively studied; however, most of these studies are either limited to the close solar neighbourhood \citep[the closest 1~kpc,][]{2010IAUS..265..304A,2009A&A...501..941H, 2007A&A...475..519H, 2007ApJ...665..767R,2004A&A...418..989N},{ they} explore in more detail some specific Galactic directions \citep{2002ApJ...574L..39G,Kordo2011b}, or they are devoted to the study of the Galactic thick disc \citep{2009IAUS..258...23F,2008A&A...480..753V,2008A&A...480...91S,2003A&A...398..141S,2003A&A...399..531S}. These different studies have improved our knowledge of the different Galactic components in the solar vicinity but we are far from completely understanding them  {over the whole Galaxy}.

The advent of Gaia \citep{2005ASPC..338...15M} will strongly constrain the structure and composition of the Milky Way because kinematical and chemical properties will be measured for millions of stars.
Before Gaia, several large-scale surveys,
 {  such as RAVE \citep{Stein} and the Gaia-ESO Survey  \citep{Gilmore12}}, are collecting data about the Galaxy, helping us to prepare to exploit the Gaia results. 
On the other hand, the \corot\ (Convection Rotation and planetary Transits) mission is collecting light curves for several thousand stars close to the Galactic plane  towards two diametrically opposed directions. Therefore, a non-negligible by-product of this mission is the study of the Galactic  structure in the directions observed by \corot. 

To prepare and support this mission, massive spectroscopic observations have been performed, resulting in good precision radial velocities for 1534 \corot~targets \citep{Loeillet2008FLAMES}. Furthermore, atmospheric parameters,  { 
\textit{i.e.}  effective temperature ( \teff), surface gravity (\logg), global metallicity (\met), and $\alpha-$elements abundances (\alf), }have been determined with the \matisse~algorithm for 1227 \corot~targets in three of the directions observed by the satellite \citep{Gazzano2010}.
 These data represent a good opportunity to identify and characterise the different stellar populations composing these Corot Fields, to explore the Galactic structure and chemistry in these directions, 
and to explore radial metallicity and abundance gradients. 

\defcitealias{Gazzano2010}{Paper I}

The previous analysis by \citet[][hereafter \citetalias{Gazzano2010}]{Gazzano2010}  aimed at demonstrating the ability to perform robust automated spectral classification.
In the present paper, we extend the characterisation of the stellar populations in three of the \corot~fields by combining kinematical and atmospheric parameters.
In Sect.~\ref{sec:sample}, we recall some properties of the three observed samples and present the kinematics analysis.
In Sect.~\ref{sec:comp}, we compare and validate our results with the Besan\c con Galactic model \citep{2003A&A...409..523R}. In Sect.~\ref{Sec:stelPop}, we separate and discuss the various stellar populations identified in our spectroscopic sample. 
{  The properties of the thin disc are analysed in Sect.~\ref{sec:thindisc}.} We also suspect the presence of a peculiar population with high and non-standard \alf~values and rather high metallicities, which is presented in Sect.~\ref{sec:highalf}. The impacts of our study in terms of planet population in these \corot~fields are discussed in Sect.~\ref{sec:planets}. Lastly, we give our conclusions in Sect.~\ref{SecDis}.

\section{Stellar properties in the targeted Galactic directions}
\label{sec:sample}

\subsection{Galactic directions studied}
We used the samples of \corot/Exoplanet targets analysed in \citetalias{Gazzano2010}. These stars are located in three of the \corot/Exoplanet fields, namely the \textit{Long Run Anticentre 01 (LRa01)}, the \textit{Long Run Centre 01 (LRc01)}, and the \textit{Short Run Centre 01 (SRc01)}. These {observations} contain relatively bright stars \citep[$J<15$, 2MASS filter system, see][]{2003tmc..book.....C} located close to the Galactic plane towards the \emph{Monoceros (LRa01)} and the \emph{Aquila (LRc01 \& SRc01)} constellations. 
\begin{figure}[!h]
\centering
\includegraphics[width=0.4\textwidth]{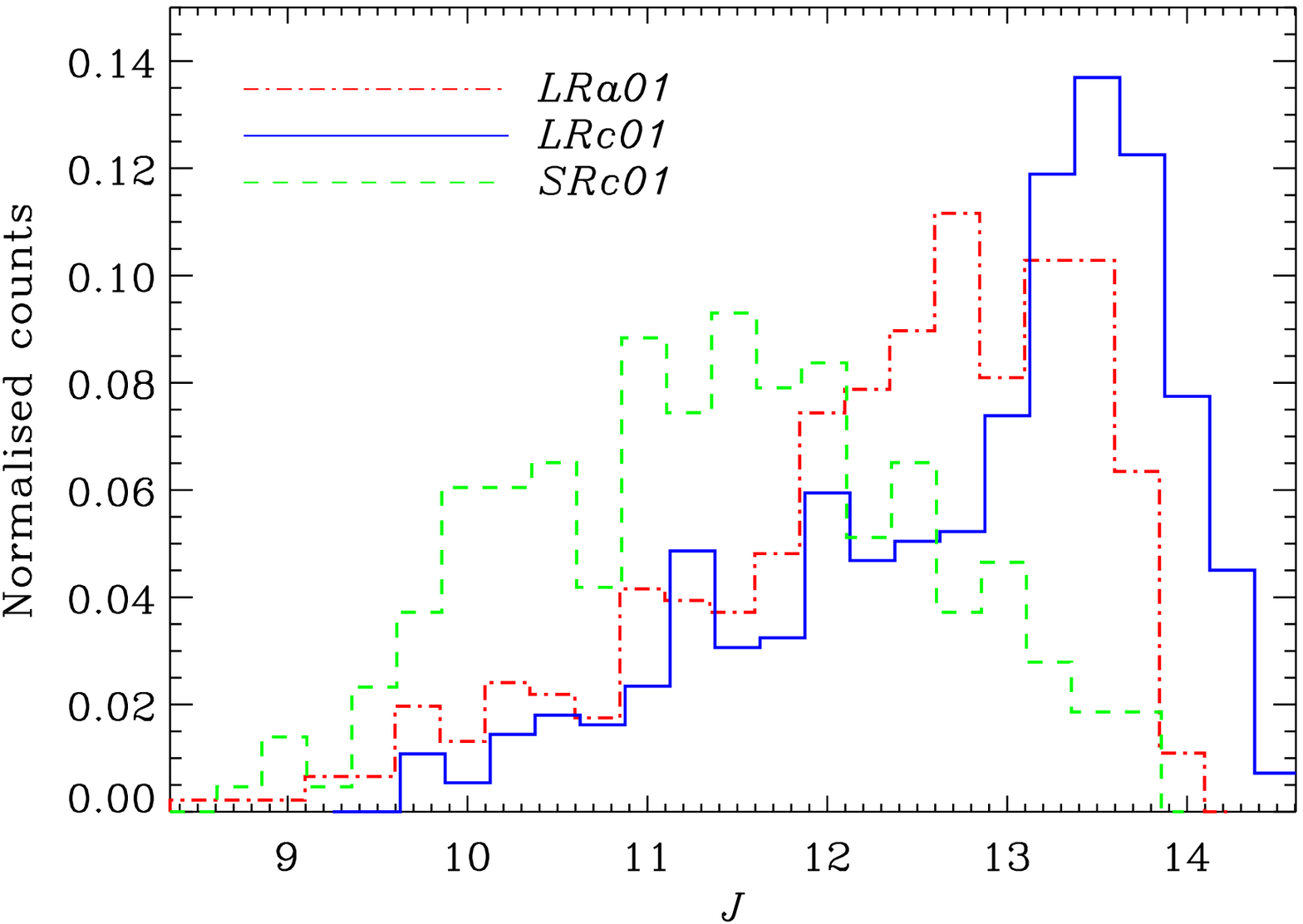}
\includegraphics[width=0.35\textwidth,angle=90.]{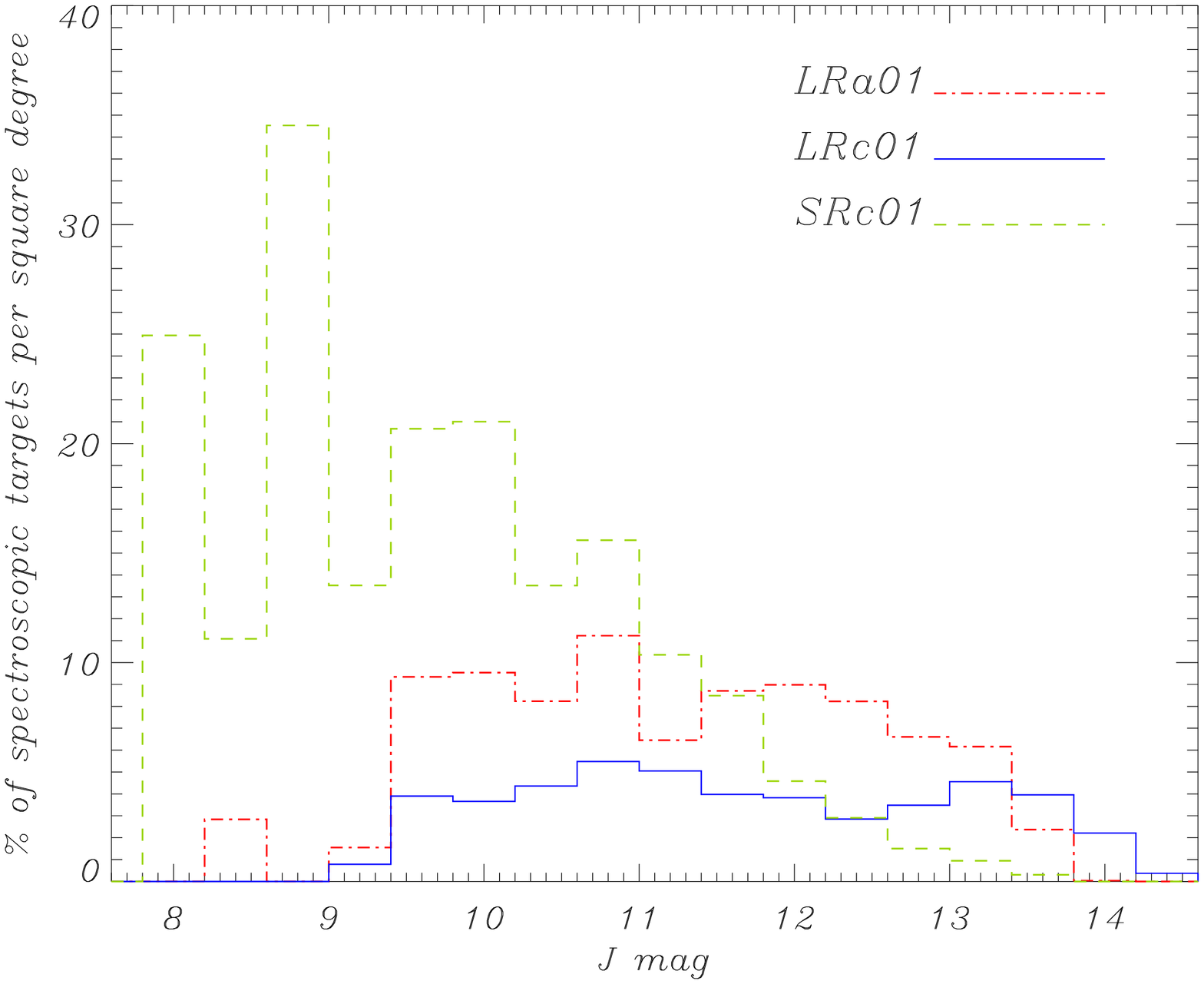}  
\caption{Properties of stars analysed in this study in each of the three pointing directions. {\it Top :} Distribution of the $J$ magnitudes of the spectroscopic sample normalised to the total number of targets ; {\it Bottom :} Distribution of the $J$ magnitudes of our spectroscopic targets given in percentage of the overall population in the corresponding \corot/exoplanet field. The red dotted-dashed represents \emph{LRa01} stars, the blue solid line stands for \emph{LRc01} stars, and \emph{SRc01} is shown with a long-dashed green line.}
             \label{fig:magr}
\end{figure}

As presented in \citetalias{Gazzano2010}, the targets were observed during two different campaigns with slightly different observation strategies. The first campaign took place in January 2005, which is before the launch of \corot, while the second one was done more than three years later, when \corot\ was in operation. In both cases, the observations aimed at both performing accurate Doppler measurements and checking and improving the spectral classification of stars in the \corot\ exoplanet fields. The spectral classification of all the stars within the range of magnitudes observable by \corot\ was mandatory to identify dwarfs and giants and to further prepare the targets to be selected for \corot\ observations in the exoplanet fields. Once the light curves are analysed, it is also useful to help the screening out of transiting systems stellar in origin. Prior to the launch, the estimate of spectral types and luminosity classes has been carried out based on photometric data from both dedicated ground-based observations and existing catalogues \citep{2009AJ....138..649D}. This allowed to build the \corot\ input catalogue, \emph{Exo-Dat}, which provides for all stars within the potential exoplanet fields of the mission, the astrometry, magnitudes in different bands in the visible and near-IR, the spectral type, and luminosity class as estimated from the photometry. For our two spectroscopy campaigns, we used similar criteria to select our targets in a given \emph{FLAMES/GIRAFFE} field, based on this input catalogue. As presented in detail in \citetalias{Gazzano2010} and \citet{Loeillet2008FLAMES}, the highest priorities were given to bright F,G, K, and M-dwarfs and subgiants with a $r^{\prime}$-magnitude less than 15.0 in order to ensure that a signal-to-noise ratio decent enough could be reached. The angular separation between stars for the positioning of fibers within the field of view of the spectrograph imposes additional constrains that require not using too strict selection criteria. As a very last priority, we thus allow the selection of any type of star fainter than magnitude 15.0 in $r^{\prime}$.

The properties of our targets in the three directions are summarised in Table~\ref{tab:propSamples}. Figure~\ref{fig:magr} shows the distributions in $J$-magnitude of the observed stars and how they compare to the overall stellar population in each \corot\ field. It is clear that our sample represents a very small fraction of the total population in a given field. This is due to the limited number of \emph{FLAMES/GIRAFFE} fields we could observe in spectroscopy during the nights that were allocated to our programmes along with the observation strategy. The percentage of targets we observed depends on the field as the stellar populations differ from one region to another. While \emph{LRc01}  and \emph{LRa01} which were selected as prime fields for the \corot\ exoplanet programme, are densely populated and present an homogeneous stellar density,  the \emph{SRc01} whose selection was driven by the \corot\ asteroseismology programme, is poorly populated and very inhomogeneous, with regions of the field affected by a strong extinction. This explains why our spectroscopic targets in the \emph{SRc01} are more concentrated toward bright stars. In contrast, the percentage of spectroscopic targets in \emph{LRc01} and \emph{LRa01} is distributed better over the magnitude range, the lowest percentage for \emph{LRc01} being due to a much higher stellar counts in this region, compared to  \emph{LRa01}.

\begin{table}
 { 
 \caption{Properties of the three Galactic directions studied.}
\label{tab:propSamples} 
\centering \tiny
\begin{tabular}{c c c c r r c}
\hline\hline
\noalign{\smallskip}
\corot~field & N$_{\rm Tot. Stars}$  & N$_{\rm F.S}$& l (\degre) & b (\degre)  & \multicolumn{1}{c}{$J$}  \\
\noalign{\smallskip}
\hline
\noalign{\smallskip}
\emph{LRa01} & 457 &  404 & 212.2 & $-1.9$ & [~8.7~;~13.9~] \\
\emph{LRc01} & 555 & 286 & 37.5 & $-7.5$ & [~9.7~;~14.6~]\\
\emph{SRc01} & 215 & 64 & 36.8 & $-1.2$ & [~9.0~;~13.8~]\\
Tot. Sample & 1\,227 & 754 & \multicolumn{1}{c}{-} & \multicolumn{1}{c}{-} & [~8.7~;~14.6~]\\
\noalign{\smallskip}
\hline
\end{tabular}
\tablefoot{The columns contain the \corot~field {names}, the number of stars with \matisse~atmospheric parameters, good kinematics parameters, and correct \sn~spectra {(Final Sample, see Sect.~2)}, the mean Galactic longitude and latitude, and the $J$ magnitude range.}
}
\end{table}

\subsection{Atmospheric and chemical stellar properties}
\label{Sec:atmP}
In \citetalias{Gazzano2010}, we used the \matisse~algorithm \citep{recioblancoetal2006} to derive stellar atmospheric parameters, \textit{i.e.} the effective temperature (\teff), the surface gravity (\logg), the overall metallicity (\met), and the $\alpha-$enhancement with respect to iron (\alf)\footnote{we considered O, Ne, Mg, Si, S, Ar, Ca, and Ti as $\alpha$ elements}, for 1\,227 stars from their \flames~spectra in the {HR9B  configuration.} These parameters are affected by several sources of uncertainty. The \textit{internal} uncertainty is the numerical uncertainty only due to  observational noise in the spectra and the \matisse~method. This was estimated in \citetalias{Gazzano2010} to analyse using \matisse\ a grid of interpolated theoretical spectra with various signal-to-noise ratios per pixel (\sn).
This uncertainty corresponds to the self consistency of the parameter estimation procedure and how it is affected by noise. It is different from the \textit{relative} uncertainty for which other sources of uncertainty have to be taken into account (for instance, uncertainty in the atmospheric parameters induced by radial velocity uncertainties, bad normalisation, and other purely instrumental issues).
In \citetalias{Gazzano2010}, we used the multiple observations of fifty stars to evaluate this source of uncertainty. 
Finally, by comparing our atmospheric parameters with several reference libraries, we can estimate an \textit{external} source of uncertainty.
\begin{figure}[!t]
\centering
\includegraphics[width=0.5\textwidth]{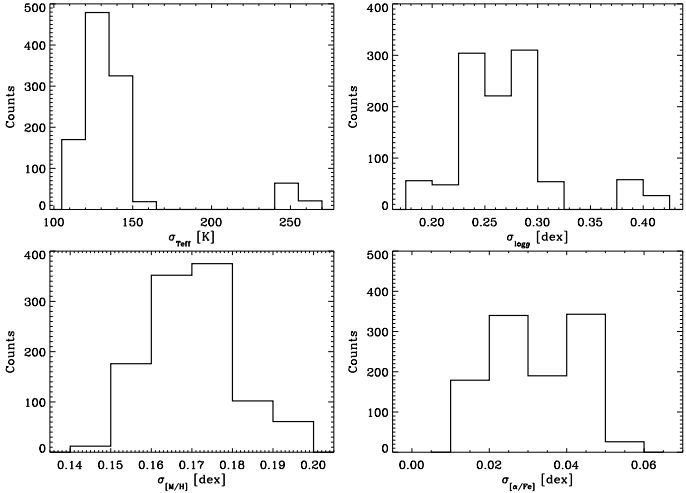}
  \caption{Distribution of the \textit{total uncertainty} for the four atmospheric parameters {for all the studied CoRoT stars}. The largest uncertainties in \teff~and \logg~distributions correspond to the few (10\%) metal-poor cool giant stars of the sample{, whose \sn~ranges from $\sim10$ to $\sim130$.}}
             \label{fig:sigMat}
\end{figure}

 {In the present study, we estimated this \textit{external} uncertainty. To that purpose, we used the same reference stars sample as in \citetalias{Gazzano2010}. This sample is composed of the 118 stars from the Elodie 3.1 library \citep{2007astro.ph..3658P}, the 90 stars from the S$^{4}$N study \citep{2004A&A...420..183A} and the 39 giant stars from the study by \citet{2009A&A...493..309S}.
We separated these reference stars sample into eight subsamples. Each of the three atmospheric parameters ranges was divided in two narrower ranges, at T$_{\rm eff}=5500$~K, \logg~=~3.5~dex, and \met~$=-0.36$~dex. 
To ensure the reliability of our uncertainty estimate, we took every source of uncertainty into  account, and calculated the \textit{total} uncertainties, adding quadratically the \textit{internal}, \textit{relative} \citepalias[both reported in][]{Gazzano2010}, and \textit{external} uncertainties.

The distributions of the \textit{total uncertainty} for the four parameters are presented in Fig.~\ref{fig:sigMat}. The largest \textit{external} uncertainties are found for the metal-poor giant cool stars (8\% of the whole sample, {79 stars with a median \sn~of 25}), for which we estimated \textit{total uncertainties} of $\sigma_{\rm T_{eff}}\simeq244$~K, $\sigma_{\log~g}\simeq0.382$~dex, $\sigma_{\rm [M/H]}\simeq0.156$~dex, and  $\sigma_{\rm [\alpha/Fe]}\simeq0.06$~dex. For the majority of the sample{, with a median \sn~of 23,} the \textit{total uncertainty} in the determination of atmospheric parameters is $\sigma_{\rm T_{eff}}\simeq125$~K, $\sigma_{\log~g}\simeq0.26$~dex, $\sigma_{\rm [M/H]} \simeq0.17$~dex, and  $\sigma_{\rm [\alpha/Fe]} \simeq0.1$~dex.
}
This approach, even if pessimistic, is the safest one because we try not to neglect any source of uncertainty. Our \textit{total} uncertainties may be overestimated because we assumed the literature parameters to be perfect whereas each measurement has associated uncertainties, and significant differences can be found between different parameter estimations performed by different authors.

\begin{figure}[!t]
\label{Fig:Georges}
\centering
\includegraphics[width=0.3\textwidth]{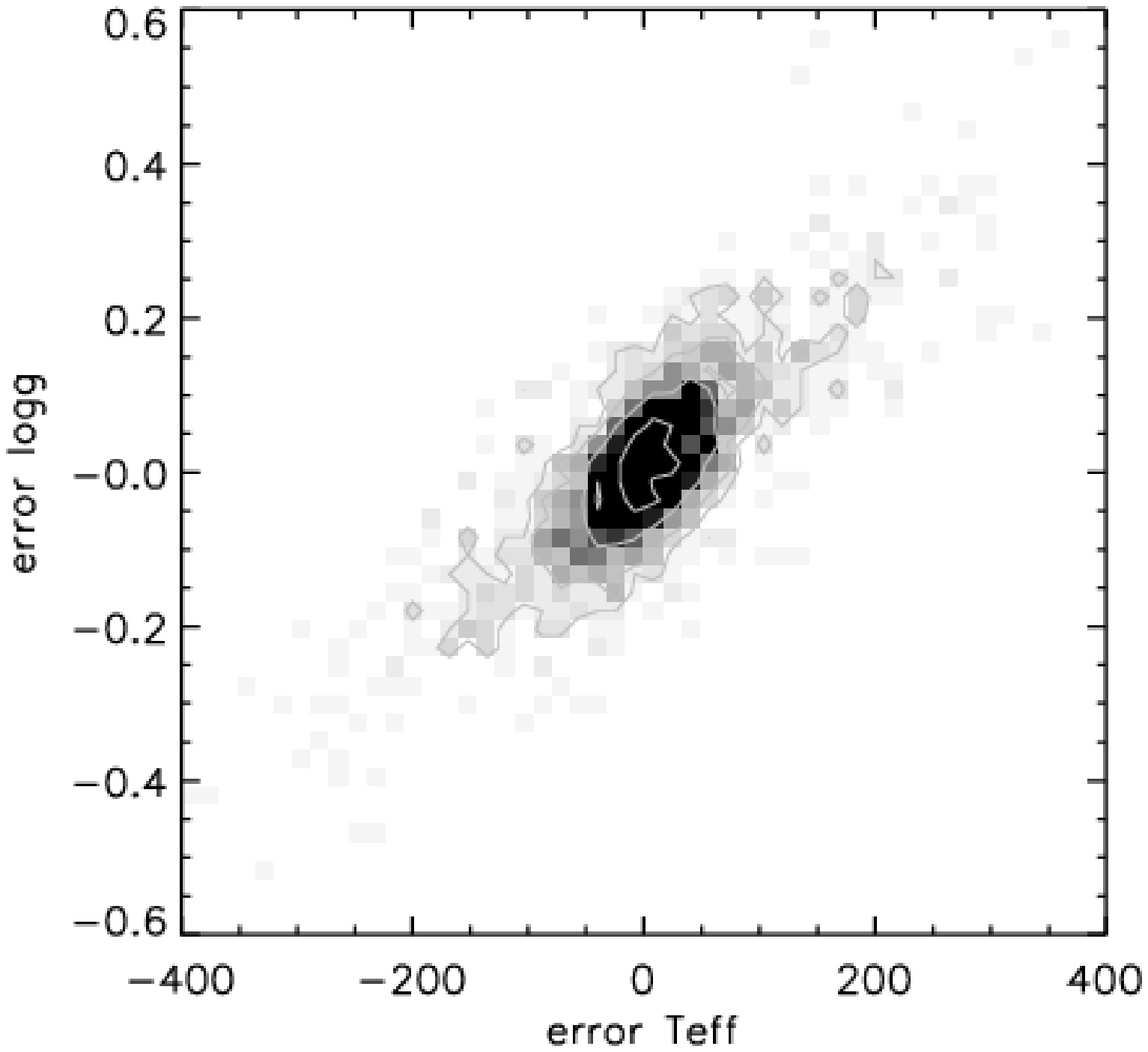}
\includegraphics[width=0.3\textwidth]{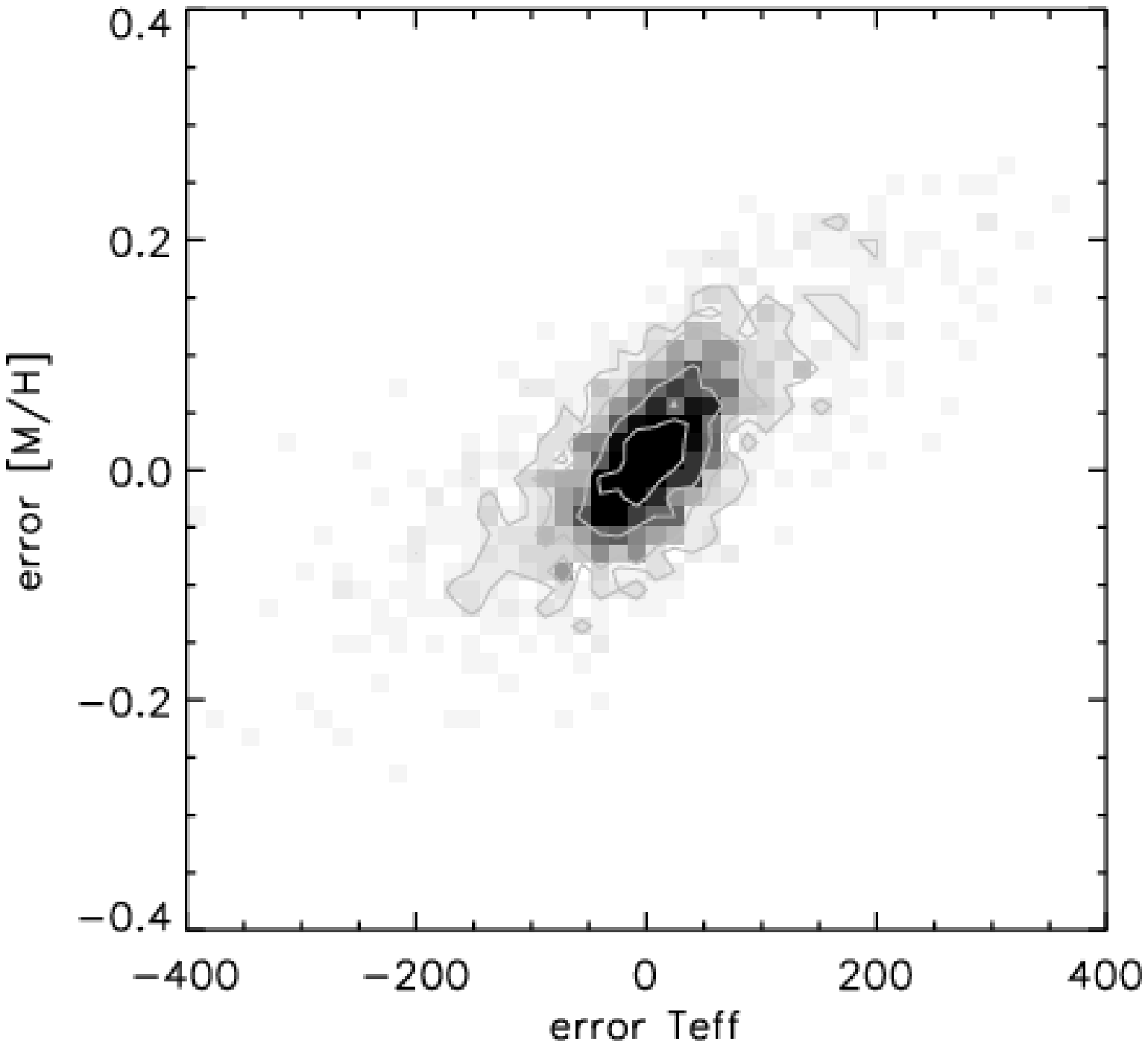}
\includegraphics[width=0.3\textwidth]{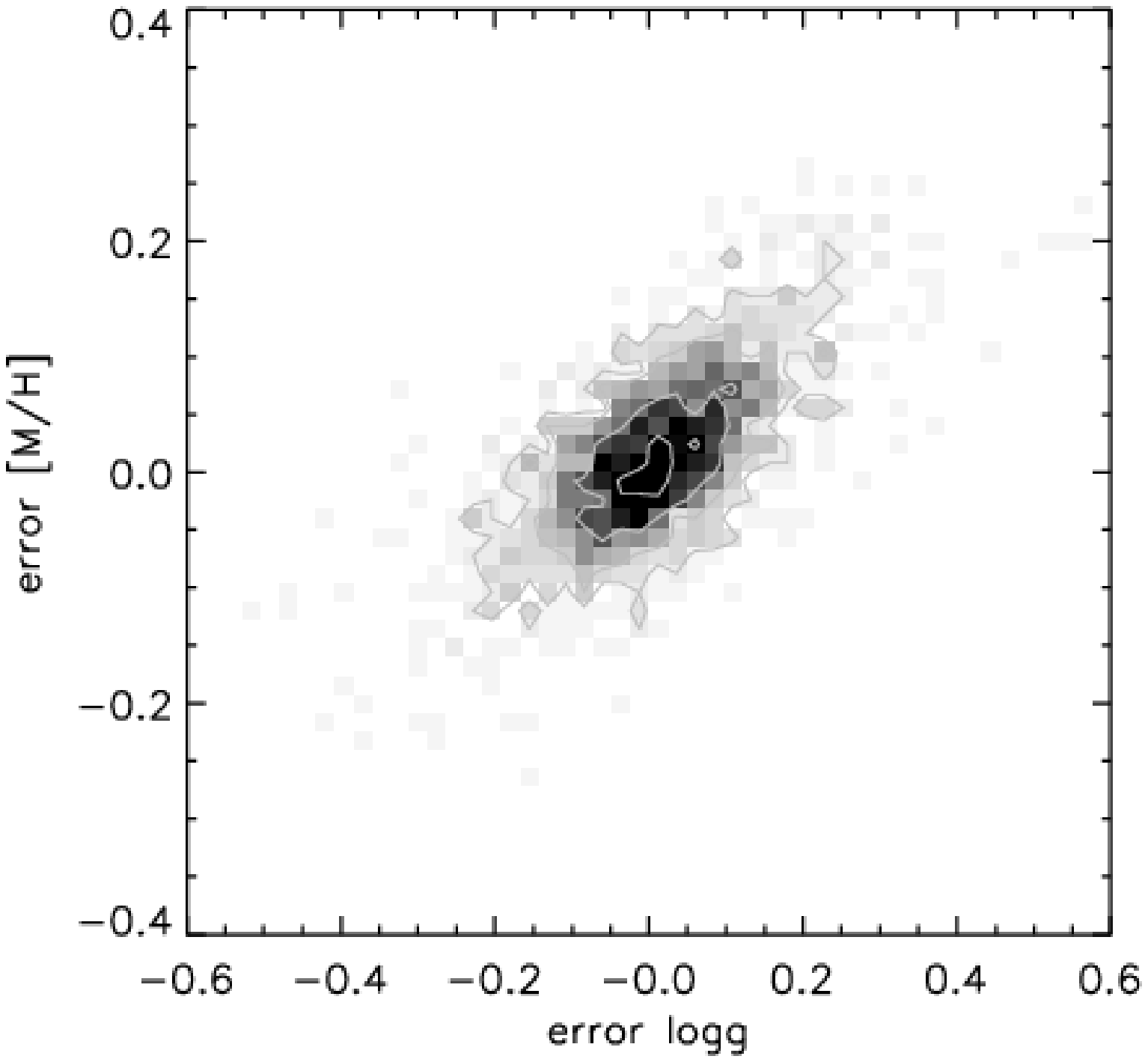}
  \caption{{   Error ellipses computed as the difference between the input and the derived parameters of our pipeline, for the set of 2,000 spectra of Sect.~\ref{sec:comp}, with SNR$\sim10$. These figures illustrate the extent of the correlation of the uncertainties in our parameter derivations. The 2-dimensional histograms have been obtained by binning by 16~K in effective temperature, 0.025~dex in $\log g$ and 0.016~dex in [M/H]. The isocontours are for 2, 5, 15 and 30 stars}}
\end{figure}

To further test the consistency between these different sources of uncertainty, we added Gaussian noise to the spectra of the S$^{4}$N sample and {re}calculated the atmospheric parameters at the signal-to-noise ratios 50, 30, 20, 10, and 5. At every signal-to-noise ratio, the S$^{4}$N values of the \alf~are recovered within the \textit{total} uncertainty. 
{   No correlation in the uncertainties  with the observed stellar parameters was noticed. As far as the theoretical correlations are concerned,  Fig.~\ref{Fig:Georges} shows the error ellipses for a set of synthetic spectra, as defined and computed in Sect.~\ref{sec:comp}, taking the total uncertainty into account. The correlations shown in this figure confirm the results already shown in \citetalias{Gazzano2010} that the error ellipses are very small. }

\subsection{{Stellar distances}}
\label{sec:dist}
To estimate the distance of the stars in our sample, we calculated their absolute magnitude from the atmospheric parameters derived by \matisse. For that purpose, we used the \textit{YYmix2} interpolation code
to generate a set of Yonsei-Yale isochrones \citep{2004ApJS..155..667D}  to derive the absolute visual magnitude ($M_{V}$) following the procedure proposed by \citet{2010A&A...522A..54Z} and implemented by \citet{Kordo2011a,Kordo2011b}.
 {   
 We note that the Yonsei-Yale isochrones do not include any evolutionary phase after the Red Giant Branch (RGB), in particular the Red Clump (RC) where some of our stars may be lying. Nevertheless, we insist that this lack does not have any influence on the derived stellar distances of our pipeline. Indeed, 
given the uncertainties in \teff \ and \logg \ (140~K and 0.27~dex,
respectively, see Paper~I), we are not able to reliably distinguish the RGB from the RC stars.  Nevertheless, we
did the exercice of selecting the stars with 4\,500$<$\teff$<$5\,000~K and
2.5$<$\logg$<3.0$ (less than 100 stars of the total sample) and modified their absolute 
J-magnitude to meet the one given by \citet{Girardi} of
M$_{\rm J}$=-0.8 mag (for solar metallicity stars). We then reprocessed the
entire analysis, and no radical changes have been noticed in the
velocity distributions or in the metallicity distributions of Table~9.
One of the reasons for that is the relatively small number of stars
concerned, and the pipeline we used roughly obtain the good
absolute magnitudes for half of these stars.  In addition, we recall
that the tests done
on synthetic data from the Besan\c{c}on model did not show any
particular distance bias for any type of star, in particular the RC stars.}
On the other hand, the interstellar absorption cannot be neglected because our pointing directions are close to the Galactic plane. To estimate this absorption, we calculated the expected unreddened colours for our sample, from the atmospheric parameters~derived by \matisse~and the \teff-2MASS colour calibration of \citet{2009A&A...497..497G}. We inverted their Eq.~(10), keeping the only physical solution.
We used the $\jmk$ colour for the extinction estimate, and \citet{1989ApJ...345..245C} calibrations to transform colour excess into absorption.
The stellar distances ($D$) were then calculated from the absolute and apparent magnitudes in the $J$ band.
\citet{2009A&A...497..497G} give a validity domain for applying their calibrations as a function of the \teff, \met, and colours.  We rejected 114 stars that do not match their validity domain.  We propagated the 1$\sigma$ uncertainties at every step of the procedure ensuring a reliable estimate of the total uncertainty on the stellar distance. The resulting  $D$, $M_{V}$, $\vmj_{0}$, $\jmk_{0}$ and $A_{J}$ are reported in Table~\ref{tab:dist} (electronic form).

\begin{table*}[!h]
\caption{Un-reddened colours, absorption, absolute magnitudes and stellar distance of the 754 \corot~stars of the present study matching our quality criteria.
\label{tab:dist}}
\centering{
\begin{tabular}{rrrrrrrrrrr}
\hline\hline
\noalign{\smallskip}
{\corot\_ID} & {$\jmk_{0}$} & {$\Delta{\jmk_{0}}$} & {$\vmj_{0}$} & {$\Delta{\vmj_{0}}$} & {$A_{J}$} &{$\Delta{A_{J}}$} &  $M_{V}$&$\Delta{M_{V}}$&{$D$ (pc)} &{$\Delta{D}$ (pc)} \\
\noalign{\smallskip}
\hline
\noalign{\smallskip}

     211652185  &   0.624   &  0.233 &    1.738  &   0.231  &   0.812   &  0.412   &  1.550  &   0.610    &   905   &    321 \\
     211652936  &   0.367  &   0.099 &    1.183  &   0.164  &   0.143  &   0.174 &    4.640  &   0.460  &     357  &      85\\
     211666039  &   0.296  &   0.096  &   1.028  &   0.145  &   0.315  &   0.183 &    3.630  &   0.690  &     511  &     171\\
  ... & ... & ... & ... & ... & ... & ... & ... & ...  & ... & ... \\
\noalign{\smallskip}
\hline
\end{tabular}}
\end{table*}

\begin{figure}[!t]
\centering
\includegraphics[width=0.3\textwidth]{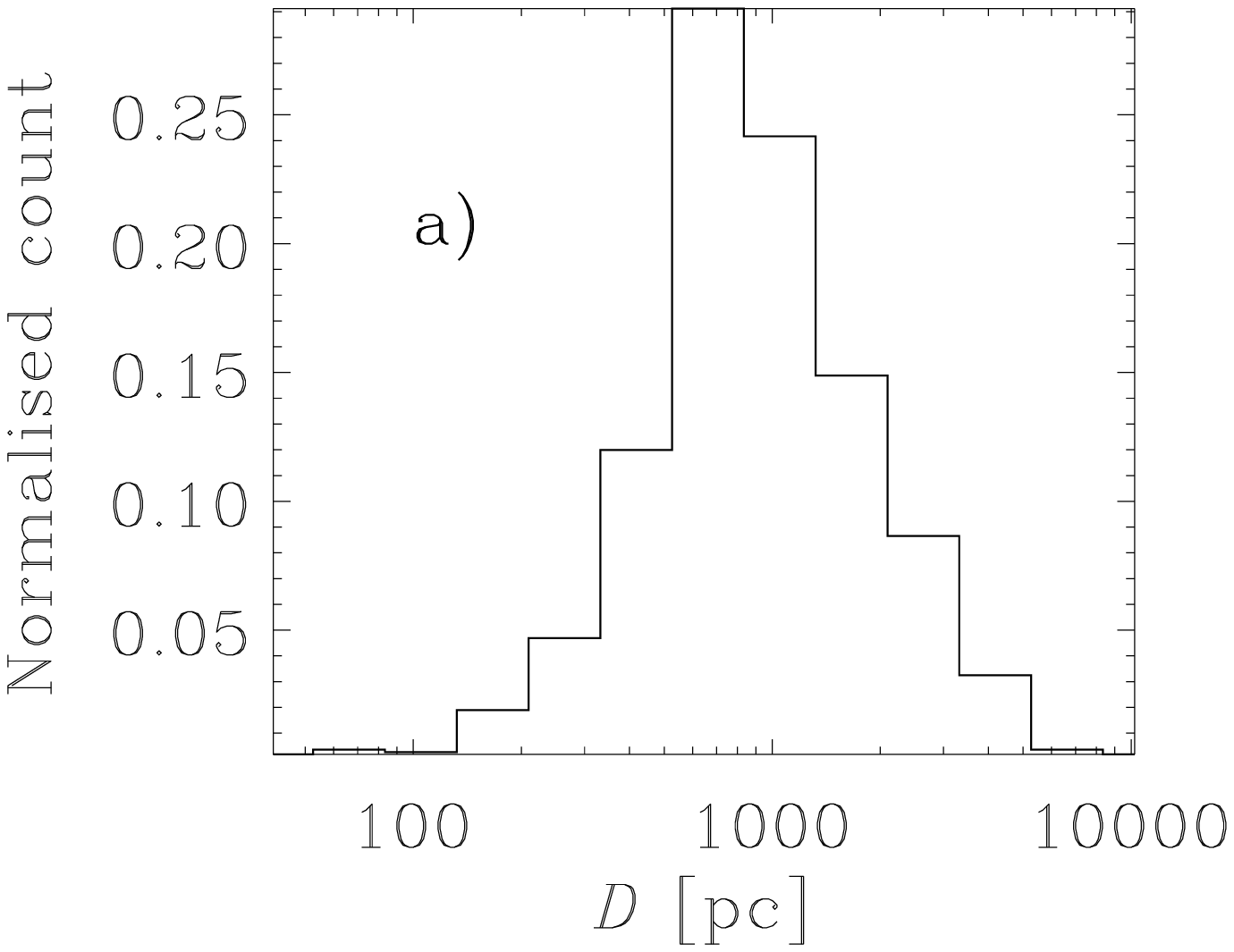}
\includegraphics[width=0.3\textwidth]{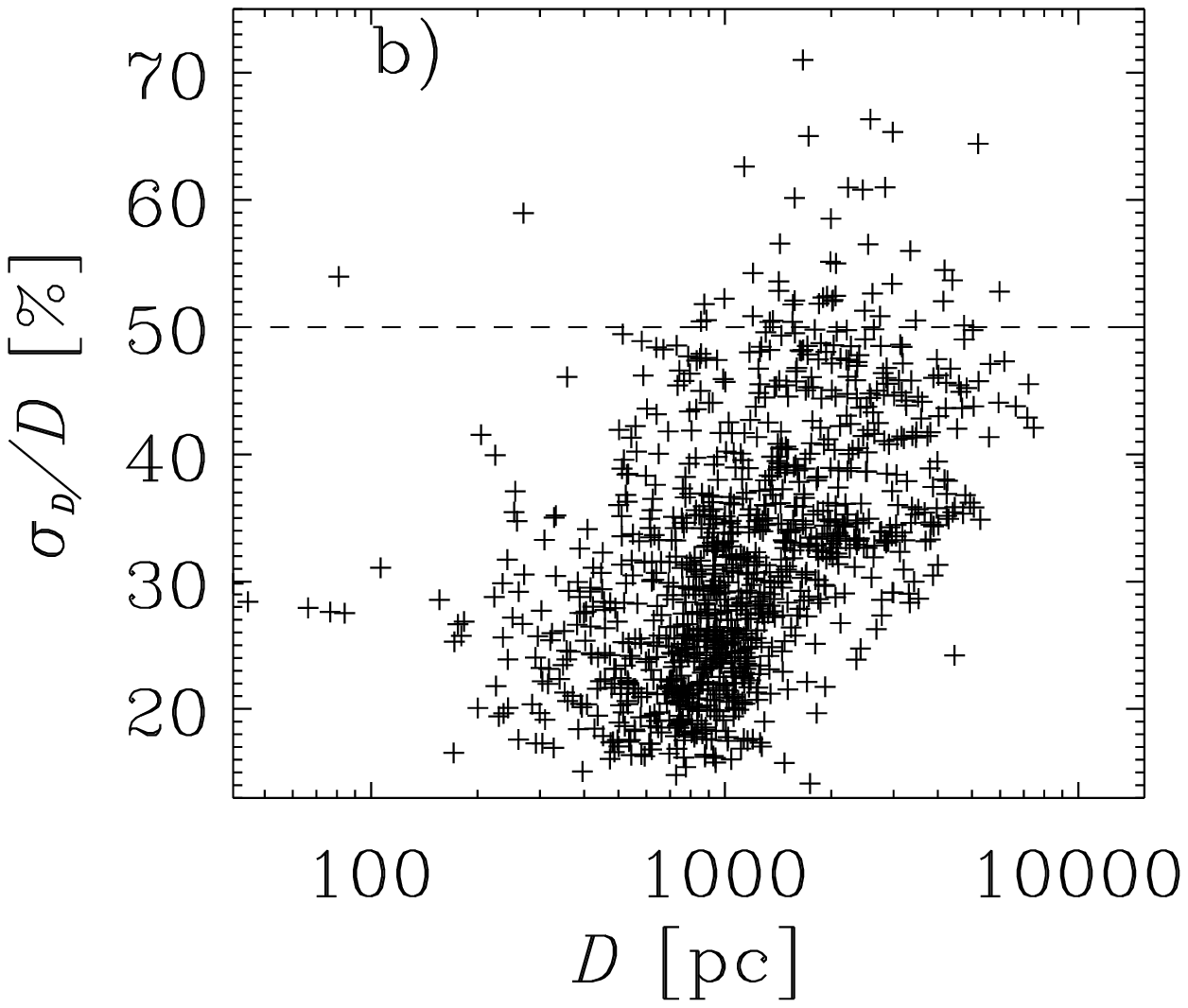}
\includegraphics[width=0.3\textwidth]{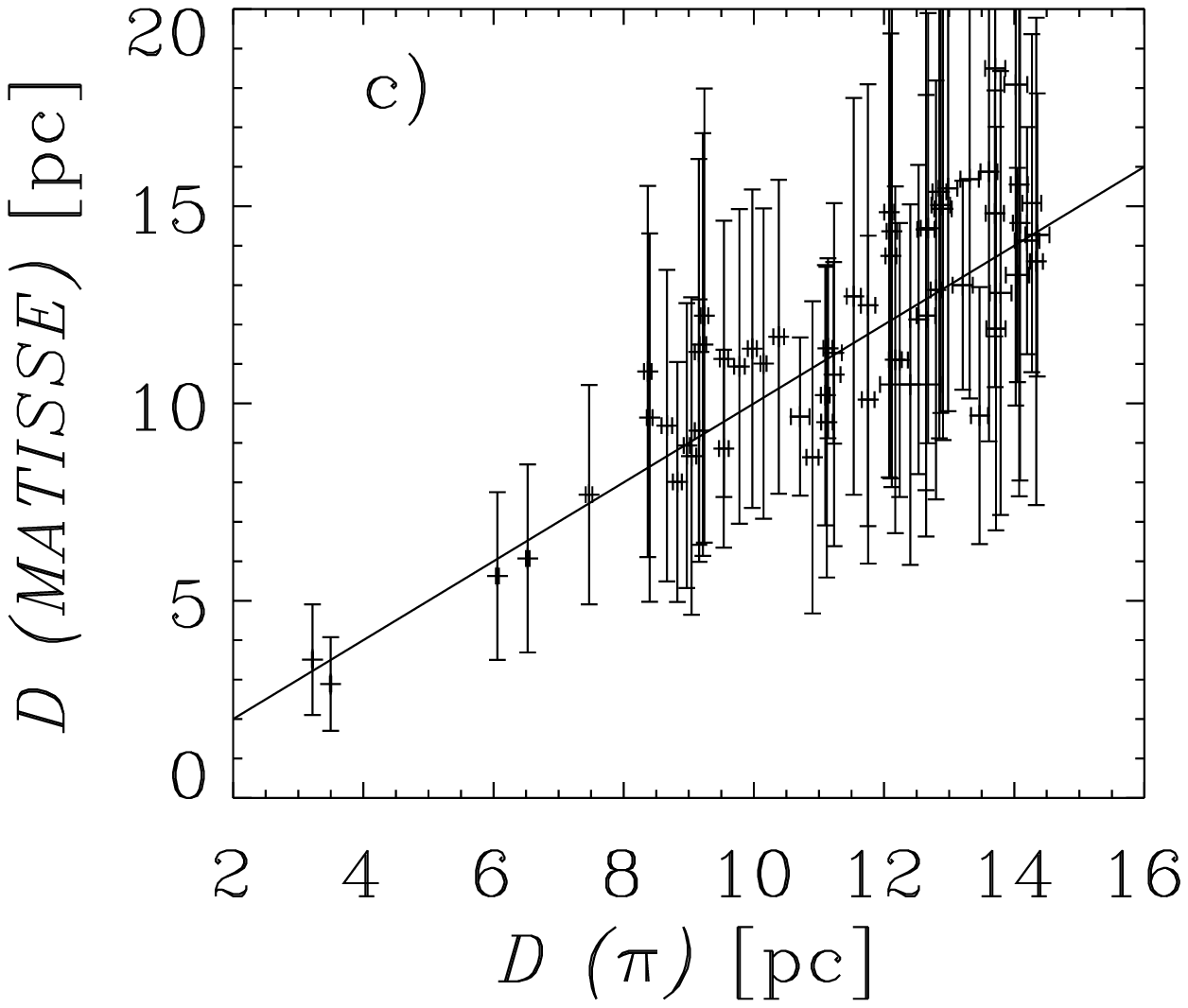}
  \caption{a) Distance distribution of the \corot~stars. b) Relative uncertainty on the distance estimation  as a function of the distance (\textit{bottom}). The dotted line represents the 50\% threshold. c) Comparison of our distance estimate with parallaxes for the S$^{4}$N sample \citep{2004A&A...420..183A}. Note that the uncertainty on the distance determination from parallaxes is about 1\% (hence not always represented). }\label{fig:sigDist}
             \label{fig:compDists}
\end{figure}

Figure~\ref{fig:sigDist}-a) shows the distribution of the stellar distance for the whole analysed sample. The vast majority is located within 2~kpc from the Sun. Note that the distribution is not symmetric, presenting a tail towards further distances, up to 6~kpc.
When dealing with distances, it is instructive to examine the relative uncertainty distribution. Figure~\ref{fig:sigDist}-b) presents the relative uncertainty on the stellar distance as a function of the distance. It shows that the typical uncertainty on the stellar distance is $\sim30$\%. Only $\sim 7$\% of the stars show uncertainty greater than 50\%, and they have  been discarded in what follows. Furthermore, we checked our distance determination by comparing it with HIPPARCOS parallaxes \citep{1997ESASP1200.....P} for 
 the stars in the S$^{4}$N sample (Fig.~\ref{fig:compDists}-b) 

 We used the atmospheric parameters given in \citetalias{Gazzano2010} and the procedure described \S\ref{Sec:atmP} for computing the \textit{total} uncertainty, assuming an \textit{internal} uncertainty of $\sigma_{\rm T_{eff}}=50$~K, $\sigma_{\log~g}=0.08$~dex, $\sigma_{\rm [M/H]} =0.05$~dex, and  $\sigma_{\rm [\alpha/Fe]}=0.02$~dex \citepalias[corresponding to \sn=20, see][]{Gazzano2010}. The correlation between the two distance determinations is 0.88, and no bias is detected ($\sigma_{\pi-D}=1.3$~pc), which validates our procedure. We therefore conclude that our stellar distance estimate is realistic.
 {   
We also note that no tests could be done on datasets of observed distant giant stars with known parallaxes, because no spectra of such stars exist in the wavelength domain and at the resolution at which we observed. Nevertheless, tests done in synthetic spectra (see Sect.~\ref{sec:comp}) have shown that we recover properly the distances for this kind of stars, even at a SNR$\sim$10.}

\subsection{{Velocity components}}
\label{sec:kine}
The kinematical data needed for the current study are the barycentric radial velocity \citep[\vrad, measured by][]{Loeillet2008FLAMES,Gazzano2010} and the proper motions ($\mu_{l},\mu_{b}$) where $l$ and $b$ are the Galactic longitude and latitude, respectively.
The proper motions were  extracted from the PPMXL catalogue \citep{2010AJ....139.2440R}. These authors showed that the UCAC3 \citep{2010AJ....139.2184Z} proper motions are {less} reliable northern than $\delta=-20$\degre, which is the case for our observations, and their uncertainties are generally low ($<\sigma_{\mu}>\simeq4$~mas yr$^{-1}$). The cross-match of the initial 1227 stars from \citet{Gazzano2010} with the PPMXL catalogue resulted in a sample of 1074 stars, \textit{i.e.} $\sim$87\% of the initial sample described in Table~\ref{tab:propSamples}. We removed all the stars with {high} uncertainties on the proper motion ($\sigma_{\mu}\ge10$ mas yr$^{-1}$) and, following \citet{2010AJ....139.2440R} recommendations, we also removed very high proper motion stars ($\mu_{\alpha}$ or $\mu_{\delta} \ge$ 150 mas yr$^{-1}$) and stars having too few observations in the PPMXL catalogue ($\rm n_{obs}\le 3$). We also removed from our sample all stars with an uncertainty in radial velocity greater than 1~\kms~(the average uncertainty being $<\sigma_{V_{\rm rad}}>\simeq 0.3$~\kms).  {   This criterium on the radial velocity is important not only in order to get small uncertainties on the derived heliocentric velocities (see below), but also to derive reliable atmospheric parameters from the stellar spectra, hence good estimations of the distances. Indeed, as shown in \citet{Gazzano2010}, for errors larger than half a pixel on the Doppler correction of the spectra, the derivation of the stellar atmospheric parameters degrades.

Finally, to interpret our results and characterise the stellar populations in these Galactic directions, we need to clean up our sample by introducing a cut in SNR. For that purpose, 
   we did not consider all the spectra with \sn~$<10$ in the following because their atmospheric parameters might be less reliable, and because for these amounts of noise the correlation in errors might become considerable. This SNR cut removes  25\% of the total initial sample of 1227 stars.
  Combined with the previously cited quality criteria, the final sample results in  
 754 stars corresponding to $\sim 62\%$ of the initial sample, \textit{i.e.} 404 stars in \emph{LRa01}, 286 stars in \emph{LRc01}, and 64 stars in \emph{SRc01} (see Table~1).  \\

The combination of stellar distances and coordinates  {allows us} to calculate the Cartesian coordinates of our sample stars whose origin is at the centre of the Galaxy:

\begin{eqnarray}
X_{GC} &=& X_{\sun}-X= X_{\sun}-D\cos{b}\cos{l}\\
Y_{GC} &=& Y_{\sun}-Y=Y_{\sun}-D\cos{b}\sin{l}\\
Z_{GC} &=& Z_{\sun}-Z=Z_{\sun}+D\sin{b}
\end{eqnarray}
where $X_{\sun}\simeq8.5$~kpc,  $Y_{\sun}\simeq0$~kpc, $Z_{\sun}=15$~pc, and ($X,Y,Z$) are the heliocentric coordinates. We also computed the corresponding space velocity components relative to the local standard of rest
\begin{eqnarray}
U &=& V_{\rm rad} \cos{b}\cos{l}-kD\mu_{b}\cos{l}\sin{b}-kD\mu_{l}\sin{l}\cos{b} \label{eq:U}\\
V &=&V_{\rm rad}  \cos{b}\sin{l}-kD\mu_{b}\sin{l}\sin{b}+kD\mu_{l}\cos{l}\cos{b}\label{eq:V}\\
W&=&V_{\rm rad}  \sin{b} + kD\mu_{b}\cos{b}\label{eq:W}
\end{eqnarray}
where $k = 4.74047$~\kms, and $(\mu_{l},\mu_{b})$ are the true proper motion{s}, \textit{i.e.} not projected on the sky. For that purpose, we converted the proper motions ($\mu_{\alpha},\mu_{\delta}$) of the \citet{2010AJ....139.2440R} catalogue into the Galactic coordinates system by using classical relations, $(\alpha_{p}=12^h49^m,\delta_{p}=27.4$\degre) as the equatorial coordinates of the Galactic pole and $l_{0}=123$\degre~as the origin of longitudes.

We remind the reader that in our convention the $U$ velocity is directed towards the Galactic centre, the $V$ towards the Galactic rotation {direction}, and the $W$ up towards the Galactic {north pole}. We propagated the 1$\sigma$ uncertainties estimated in the previous sections on the space velocity components and the Galactocentric coordinates. Clearly, the dominant source of uncertainty is the stellar distance, compared to the radial velocity and the proper motions. 
All these quantities are presented in Table~\ref{tab:kino}.

\begin{table*}[!h]
\caption{Kinematics results for the 754 stars fulfilling our quality criteria.\label{tab:kino}}
\tiny{
\begin{tabular}{rrrrrrrrrrrrrrr}
\hline\hline
\noalign{\smallskip}
  \multicolumn{1}{c}{CoRoT\_ID} &
  \multicolumn{1}{c}{$\mu_{l}$} &
  \multicolumn{1}{c}{$\mu_{b}$} &
  \multicolumn{1}{c}{$X$} &
  \multicolumn{1}{c}{$\Delta X$} &
  \multicolumn{1}{c}{$Y$} &
  \multicolumn{1}{c}{$\Delta Y$} &
  \multicolumn{1}{c}{$Z$} &
  \multicolumn{1}{c}{$\Delta Z$} &
  \multicolumn{1}{c}{$U$} &
  \multicolumn{1}{c}{$\Delta U$} &
  \multicolumn{1}{c}{$V$} &
  \multicolumn{1}{c}{$\Delta V$} &
  \multicolumn{1}{c}{$W$} &
  \multicolumn{1}{c}{$\Delta W$} \\  
  &
  \multicolumn{2}{c}{(mas cent$^{-1}$)} &
  \multicolumn{6}{c}{(kpc)} &
  \multicolumn{6}{c}{(\kms)} \\
\noalign{\smallskip}
\hline
\noalign{\smallskip}

  211652185  &  -0.27413   & -1.24966  &  -7.781   &  0.255  &   0.551   &  0.196 &   -0.017   &  0.006   &   -3.3   &   12.3   &   -4.0    &  16.0   &   -5.3   &   20.3 \\
     211652936  &  13.30323   &  4.81142 &   -8.217   &  0.068  &   0.218  &   0.052 &   -0.007   &  0.002  &   -34.0   &    5.1  &     2.3    &   6.7   &    8.6   &    6.7 \\
     211666039   &-17.94455 &   -7.38882  &  -8.088 &    0.138  &   0.302 &    0.101  &  -0.017 &    0.006   &   46.8  &    10.0  &   -19.6  &    13.9 &    -18.8  &    10.8 \\
 ... & ...& ... & ... & ... & ... & ... & ...& ... & ... & ... & ... & ... & ... & ...  \\
\noalign{\smallskip}
\hline
\end{tabular}}
\end{table*}

\section{Comparisons with the Besan\c con Galactic model}
\label{sec:comp}

We used the Besan\c con Galactic model \citep[hereafter BGM, see ][]{2003A&A...409..523R} to simulate the three pointing directions described in Sect.~\ref{sec:sample}. 
 These simulations allowed us to test our observations in the context of the canonical scenario for Galactic structure and chemistry.
 Besides, by comparing our observables to the simulation results, we can check that the observational selection biases are correctly taken into account.

For each given field, we performed a BGM request with the faint limit in $J$ magnitude, the mean Galactic longitude and latitude, a solid angle of five square degrees, which is compatible with the size of each {\flames~pointing direction}, and no extinction law. Each request provides a sample of simulated stars with their intrinsic properties (absolute magnitude, effective temperature, gravity, age, metallicity [Fe/H], U,V,W velocities computed without uncertainties), and the corresponding observables (apparent magnitude, colours, proper motion, radial velocity, distance to the Sun, and interstellar extinction).

\begin{figure}[!b]
\centering
\includegraphics[width=0.45\textwidth]{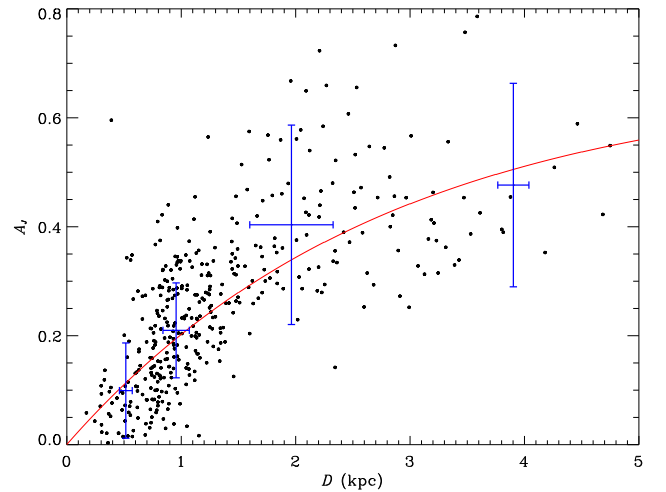}
\caption{ {Derived extinction law (red line) from the calculated absorptions and distances in} the \emph{LRa01} direction. {The typical error bar is represented for 500 pc, 1 kpc, 2 kpc, and 4 kpc.}}
             \label{fig:fitExt}
\end{figure}

The extinction being poorly constrained in the Galactic plane, we decided to apply \textit{a posteriori} {to the BGM stars} an extinction law fitted to our data. We chose to model the extinction by fitting
\begin{equation}
A_{J} = a + d \exp(cD - b) 
\end{equation}
where $D$ is the stellar distance and $A_{J}$ is the absorption in the $J$ band (both presented in Table~\ref{tab:dist}). The resulting $a$, $b$, $c$, and $d$ coefficients are given in Table~\ref{tab:fitExt} for each field. 
As boundary conditions, we forced these laws to be null at zero distance and equal to the \citet{1998ApJ...500..525S} values of the absorption at very large distances. Figure~\ref{fig:fitExt} illustrates this fit in the \emph{LRa01} direction.
\begin{table}[!h]
\caption{Adopted coefficients for the fit of the assumed extinction law applied to the BGM data and number of simulated stars in the BGM.\label{tab:fitExt}}
\centering 
\begin{tabular}{c c c c c r}
\hline\hline
\noalign{\smallskip}
& $a$ & $b$ & $c$ & $d$ & Nb BGM\\
\noalign{\smallskip}
\hline
\noalign{\smallskip}
\emph{LRa01} & 0.671 &  0.850& $-0.358$& $-1.569$ &26\,763 \\
\emph{LRc01} & 0.380  & $1.205$ & $-3.348$& $-1.155$ &100\,352 \\
\emph{SRc01} & 2.430& $-0.314$ & $-0.472$  & $-1.516$& 397\,730 \\
\noalign{\smallskip}
\hline
\end{tabular}
\end{table}

To compare our sample to the BGM, we  {   need to take } the observational selection criteria  into account by biasing each BGM request  on which we applied our extinction law.  This consists in reproducing the distributions of the infrared colour $\jmk$ and magnitude $J$ (shown in Fig.~\ref{fig:magr}).
This procedure is statistically robust for each of the three fields since the BGM requests contain a sufficient enough number of stars ($>10^{4}$, see  {Table~\ref{fig:fitExt}}).

\begin{figure*}[!T]
\centering
\includegraphics[width=0.48\textwidth]{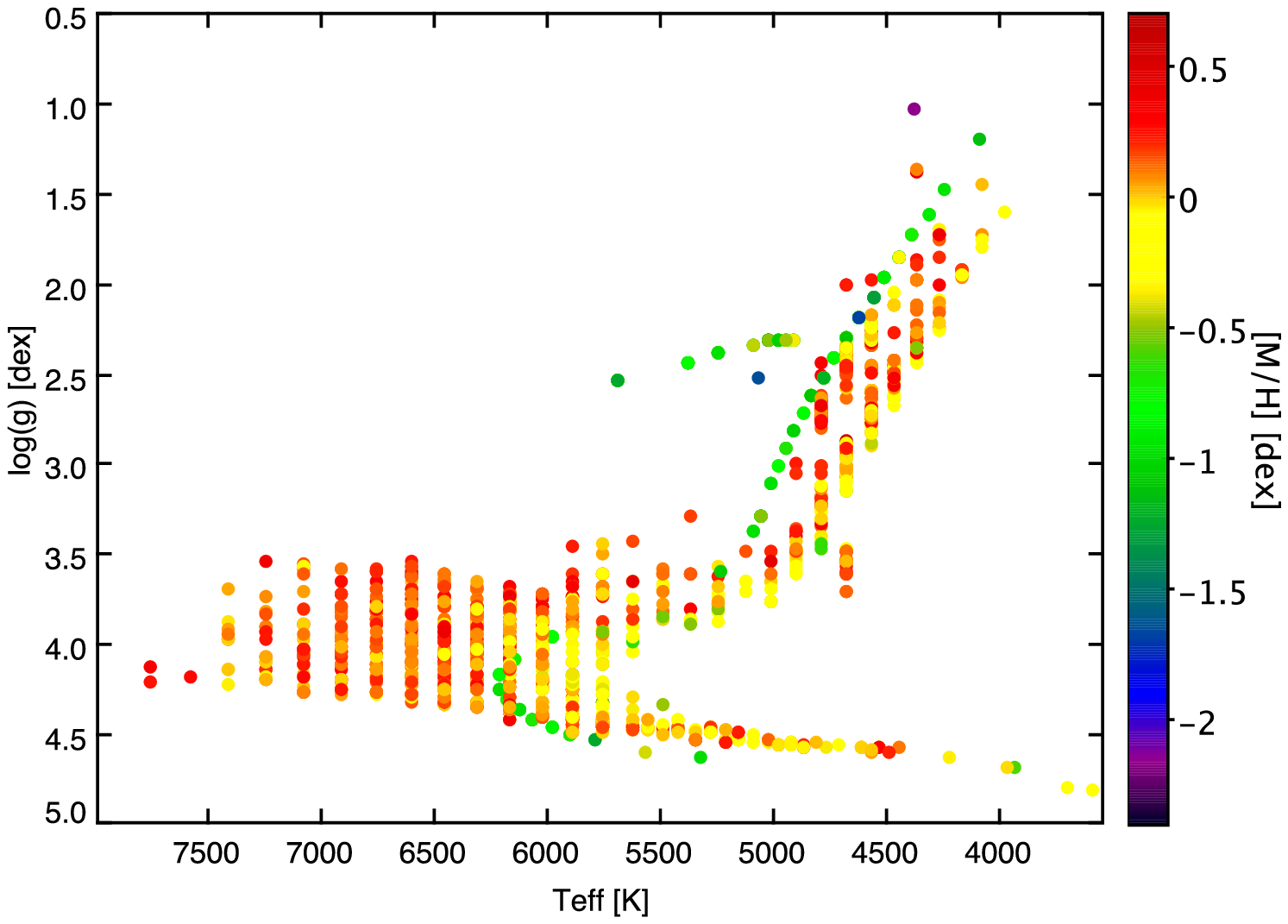}
\hfill
\includegraphics[width=0.48\textwidth]{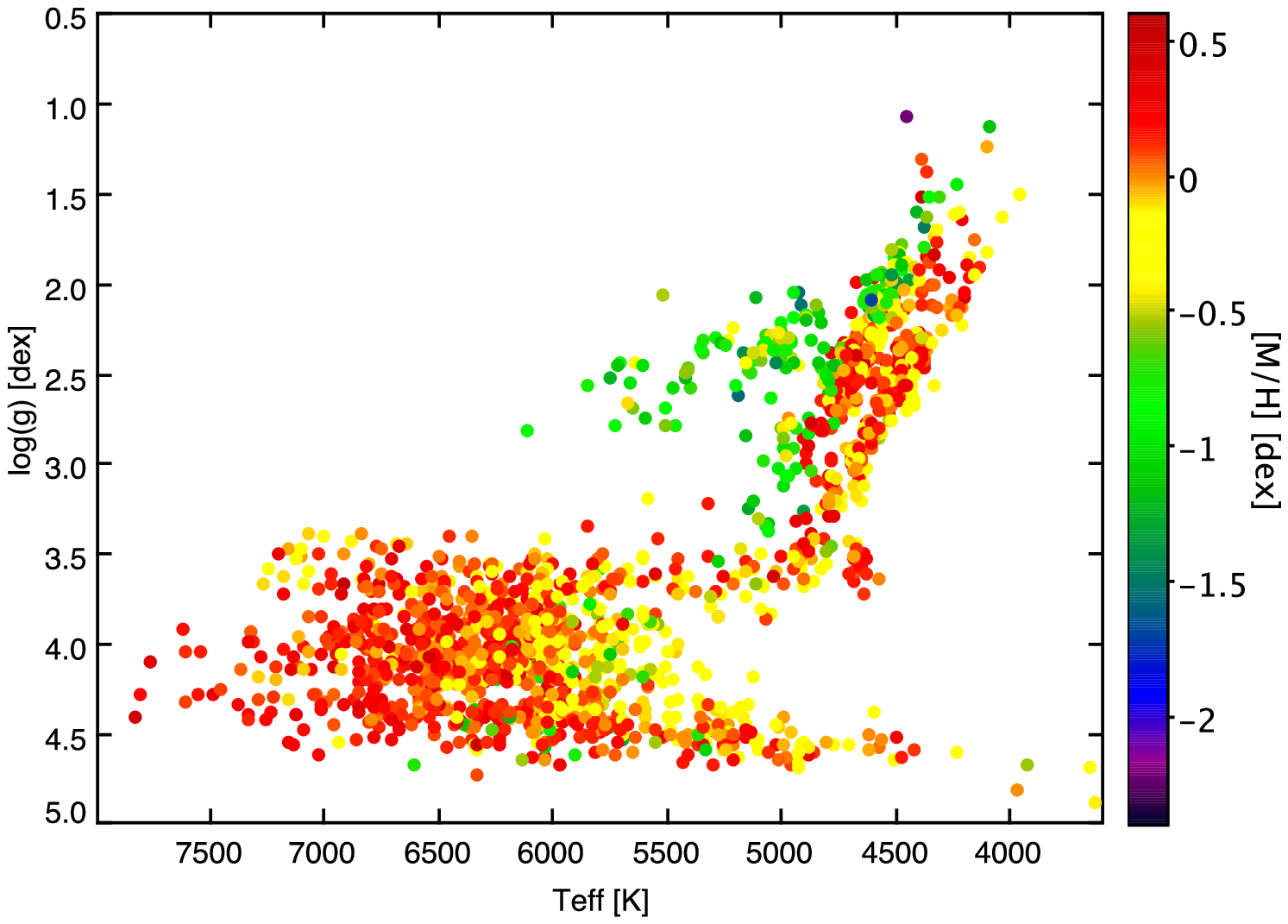}
  \caption{Hertzsprung Russel diagram in the \teff-\logg~plane for 2\,000 random entries of the BGM request in the \emph{LRc01} direction (left) and the resulting \matisse~atmospheric parameters derived from the corresponding synthetic spectra at \sn=10.}
             \label{fig:hrcomp}
\end{figure*}
\begin{figure*}[!h]
\centering
\includegraphics[angle=90,width=0.98\textwidth]{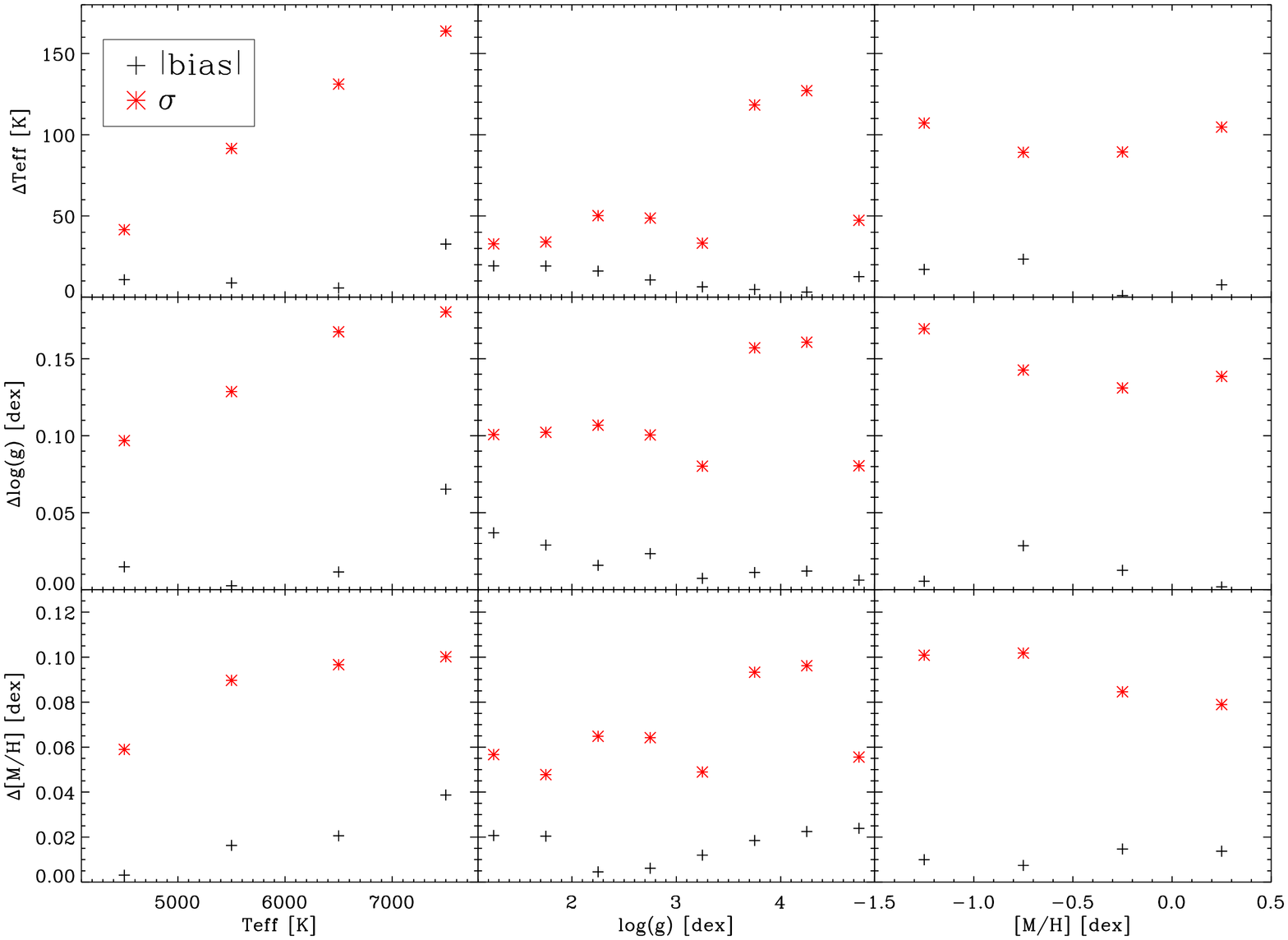}
  \caption{Evolution of the bias (black crosses) and dispersion (red asterisks) of the \matisse~atmospheric parameters for synthetic spectra (at \sn=10) of 2\,000 stars randomly chosen in the \emph{LRc01} BGM request.}
             \label{fig:heter}
\end{figure*}
\subsection{Self-consistency check of the distance calculation method}
We took advantage of the BGM request in the \emph{LRc01} direction to test our stellar distance estimates, derived from the determination of atmospheric parameters, using simulated \flames~spectra.
To that purpose, we selected 2\,000 stars randomly in this simulated sample and interpolated the corresponding theoretical spectra, {from the grid used for the learning phase of \matisse~\citepalias[see][]{Gazzano2010}, hence} with the same observational setting (wavelength range
and resolving power) as our observed spectra (see~Fig.~\ref{fig:hrcomp} \textit{left} panel). 
We added gaussian noise to these synthetic spectra and analysed them with \matisse~in the same way as our observations \citepalias{Gazzano2010}. At \sn=10, we obtained the HR diagram presented in the right hand part of Fig.~\ref{fig:hrcomp}, which is not distorted but only scattered. The agreement between the input and recovered diagrams is very good.

To study the uncertainty on the retrieved stellar parameters, introduced by \matisse, we calculated the bias and standard deviation for several combinations of the stellar parameters (see Fig.~\ref{fig:heter}). The bias is systematically very small and lower than the dispersion for any type of  stars. We found values of the \emph{internal} uncertainty that are compatible with \citetalias{Gazzano2010}. The standard deviation of the three parameters depends strongly on the effective temperature and the metallicity: cool and metal-rich stars present lower $\sigma$ values than hot and metal-poor stars.
 
Applying the entire procedure described in Sect.~\ref{sec:dist}, we thus conclude that our procedure does not introduce any internal bias in the determination of stellar distances, even at low SNR, as illustrated in Fig.~\ref{fig:DistvsDist}.
We also checked that, given the  {   derived  distances with our pipeline and  the BGM proper motions and  radial velocities}, we are able to perfectly recover the velocity components ($U,V,W$) with Eqs.~\ref{eq:U}$-$\ref{eq:W}. This ensures good consistency between our {results} and BGM simulations.  
{ 
 We insist that this test complements Sect.~\ref{sec:dist}, since it also shows that   our pipeline derives accurate distances for the distant giant stars and that the mild correlation in the derived atmospheric parameter uncertainties does not affect the final positions or velocities (see Sect.~\ref{sect:kinematics}).
}  

\begin{figure}[!t]
\centering
\includegraphics[width=0.45\textwidth]{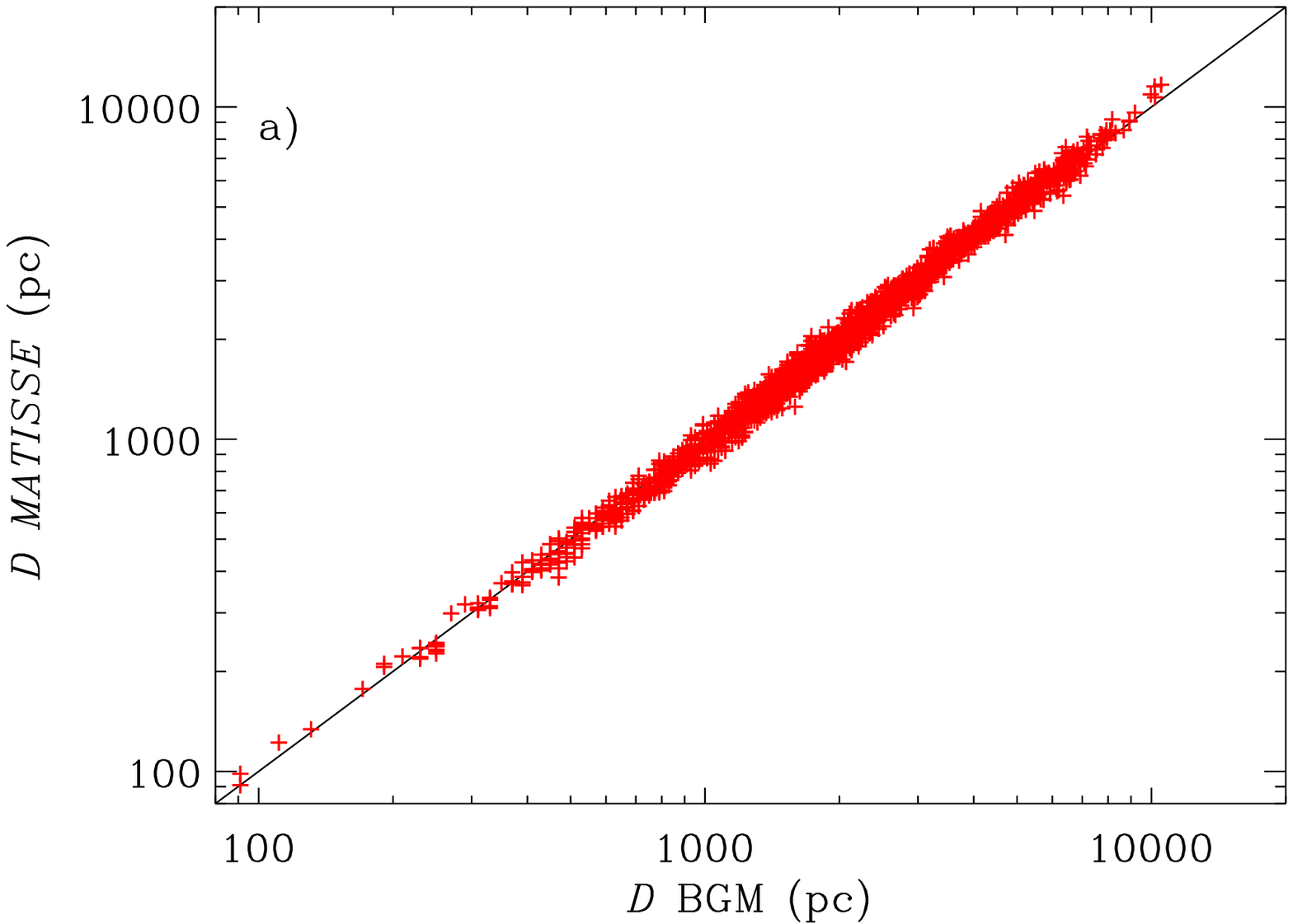}
\includegraphics[width=0.45\textwidth]{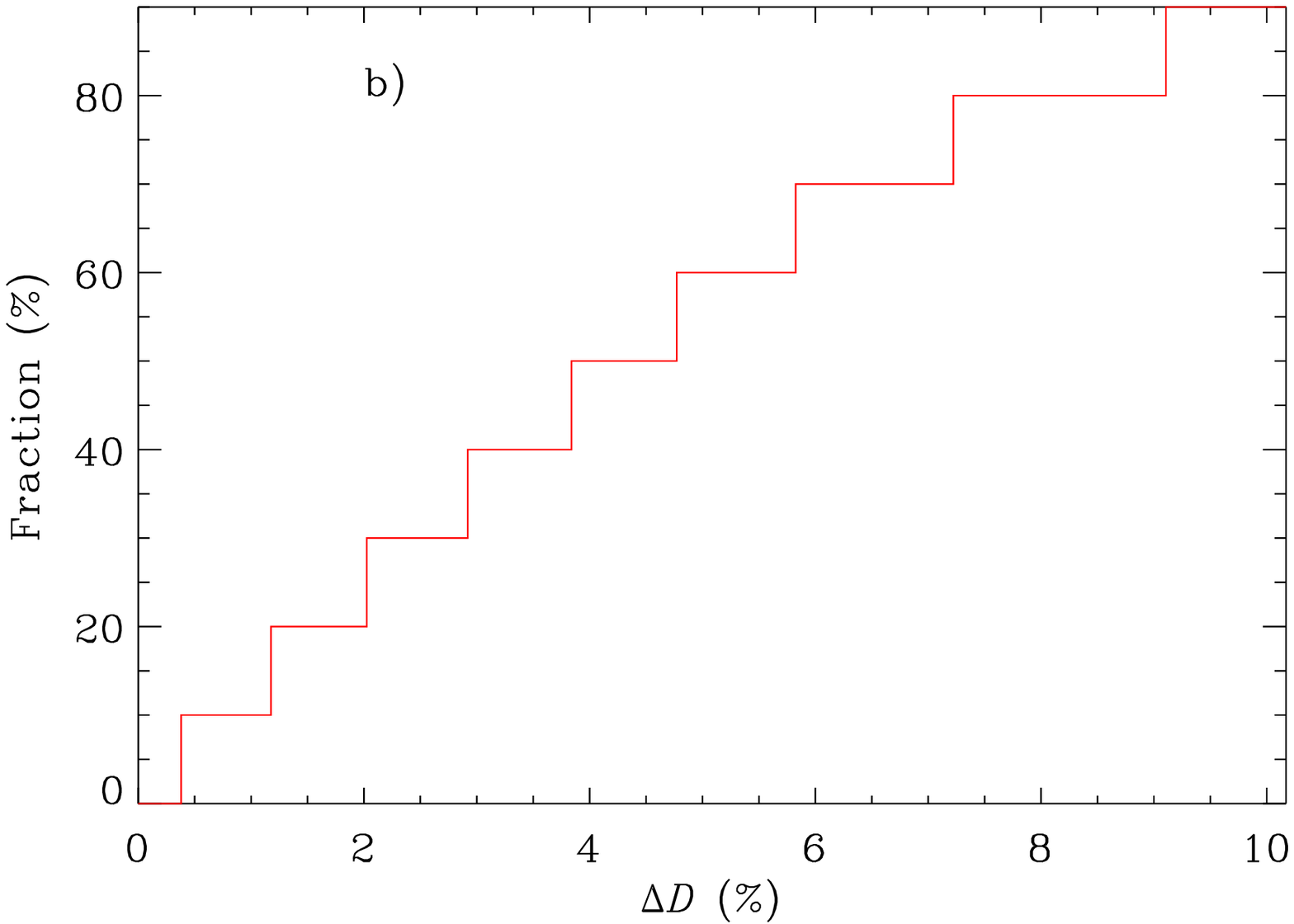}
  \caption{a)  {Derived} distances from the \matisse~parameters using the interpolated theoretical spectra having a SNR equal to 10 versus the distance computed by the BGM. b) Quantiles of the relative difference {between our stellar distances and the BGM ones.}}
\label{fig:DistvsDist}
\end{figure}

\subsection{Stellar distance distributions comparison}
Comparison of our stellar distance distributions with those predicted by the BGM is an important  {verification} to understand which regions of the Galaxy were observed and how our selection bias limits the interpretation of these data.
We, therefore, tried to estimate the effects of the parameter uncertainties on the  distance distributions. For that purpose, we considered, for every star of the sample, that its distance can be represented by a Gaussian distribution centred on the derived distance, whose $\sigma$ is the uncertainty on its stellar distance ($\sigma_{D}$). We then performed 2\,000 Monte-Carlo realisations by randomly 
picking distances in this distribution and we estimated a mean distribution, and an error bar associated to every bin.
To apply the same procedure on  the BGM simulated stellar distances, which are provided without associated uncertainties, we added,
{  for all the stars farther than 250~pc,} 
an uncertainty that was fitted on our data with a linear regression (Fig.~\ref{fig:compDists}-b)

\begin{equation}
\sigma_{D} = 0.4525D -99.32 , 
\end{equation}
which is an upper limit of the uncertainty on the stellar  distance.

The comparison of the distribution of the stellar distance showed that we also had to bias the BGM requests according to the luminosity classes, as could be expected given the sample selection criteria outlined in Sec. 2. We preferred to use the spectroscopic surface gravity since it is the parameter directly derived with \matisse, with a bin size corresponding to a luminosity class size, \textit{i.e.} 1.0~dex
\footnote{
{ 
We note that different gravity bins have been tested (0.75, 1.25, 1.5~dex) with no significant changes in the results that are presented.}}. 
We therefore constrained the BGM distributions of magnitude $J$, colour $\jmk$ and surface gravity to be identical to the observed ones. This indeed makes the distance distributions more compatible. The resulting distributions are presented in Fig.~\ref{fig:distH}. We refer to this simulated sample as BGM1 sample hereafter.
The agreement is good in all three directions, but it seems that we observed closer stars than predicted by the BGM. We should, however, point out that this assumes a correct extinction law, which is difficult to test, particularly in the \emph{SRc01} direction where differential absorption is noted.
The observations and BGM1 distance distributions agree within 2$\sigma$.

When building the metallicity distributions for our samples and the BGM one (see Fig.~\ref{fig:metaHB}), we noted that the shape and agreement with the BGM1 distributions depend strongly on the biasing in luminosity class, hence on the distance distribution of the observations. Subsequently, our metallicity distributions agree with the BGM1 distributions within 3$\sigma$ and this is strongly correlated with the distance distribution.
We conclude from these comparisons that it is indeed mandatory to simulate properly the selection biases according to the luminosity classes, since they have a direct impact on the observed distance.

\begin{figure*}[!t]
\centering
\includegraphics[width=0.33\textwidth]{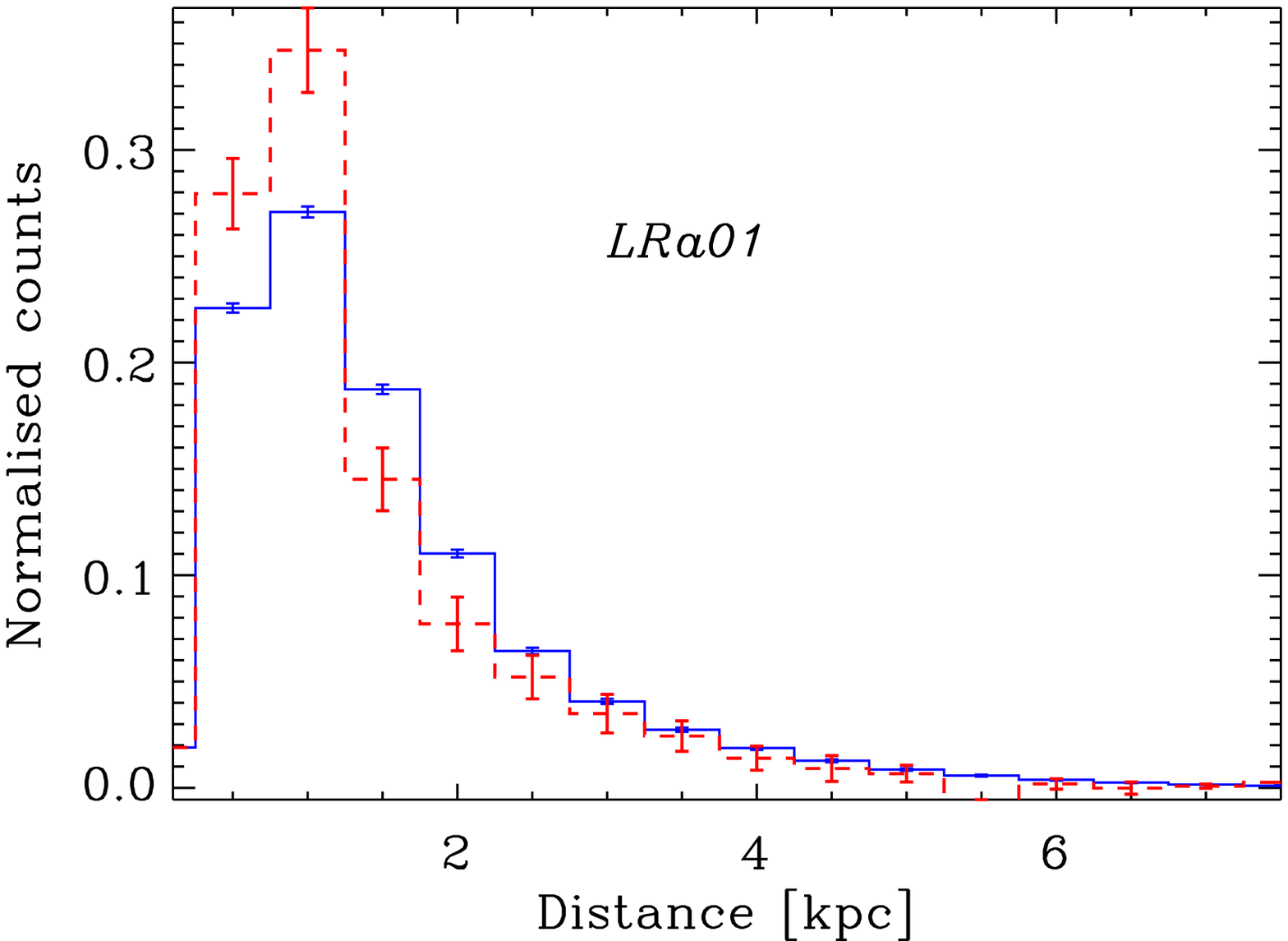}
\includegraphics[width=0.33\textwidth]{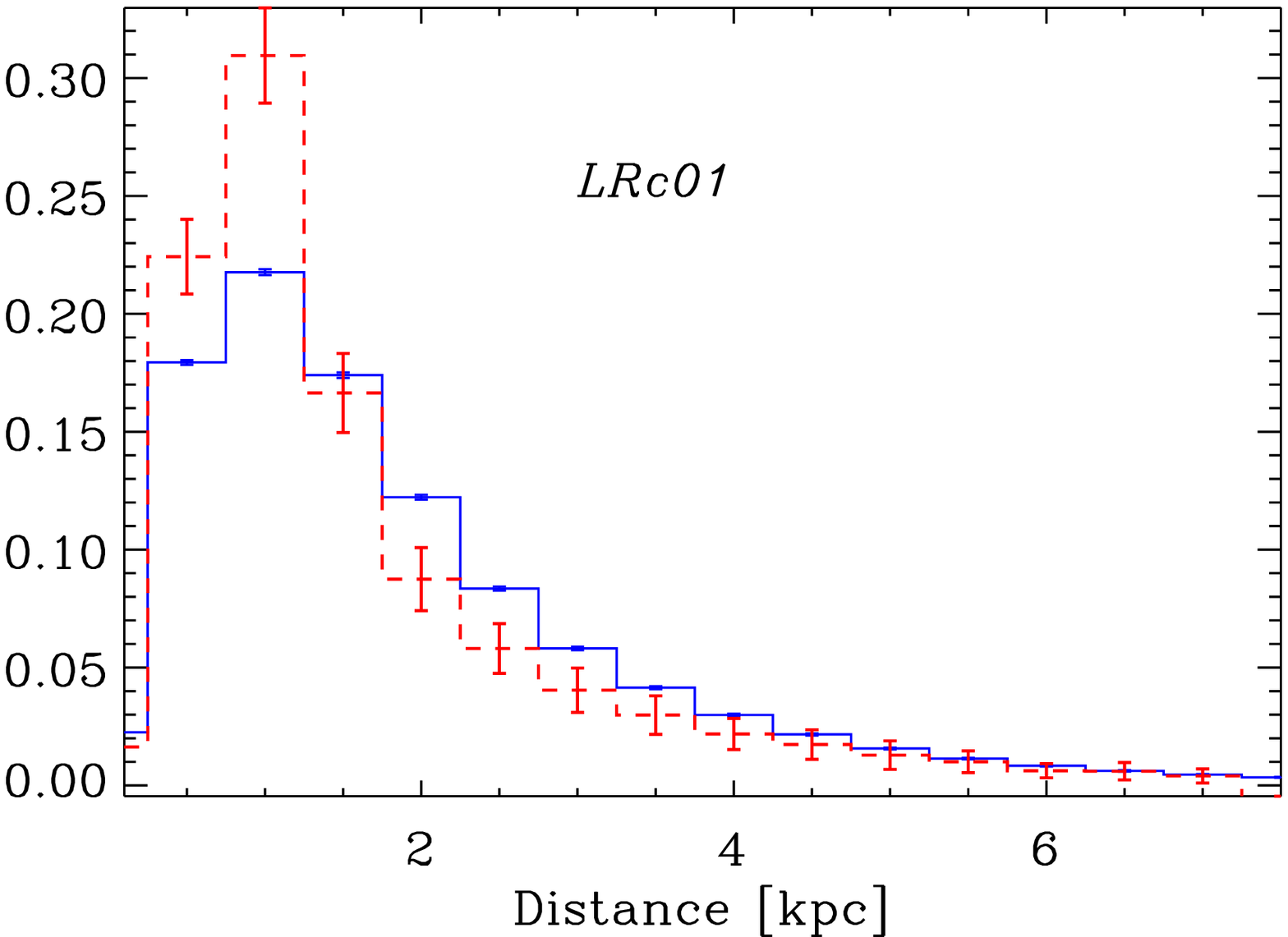}
\includegraphics[width=0.33\textwidth]{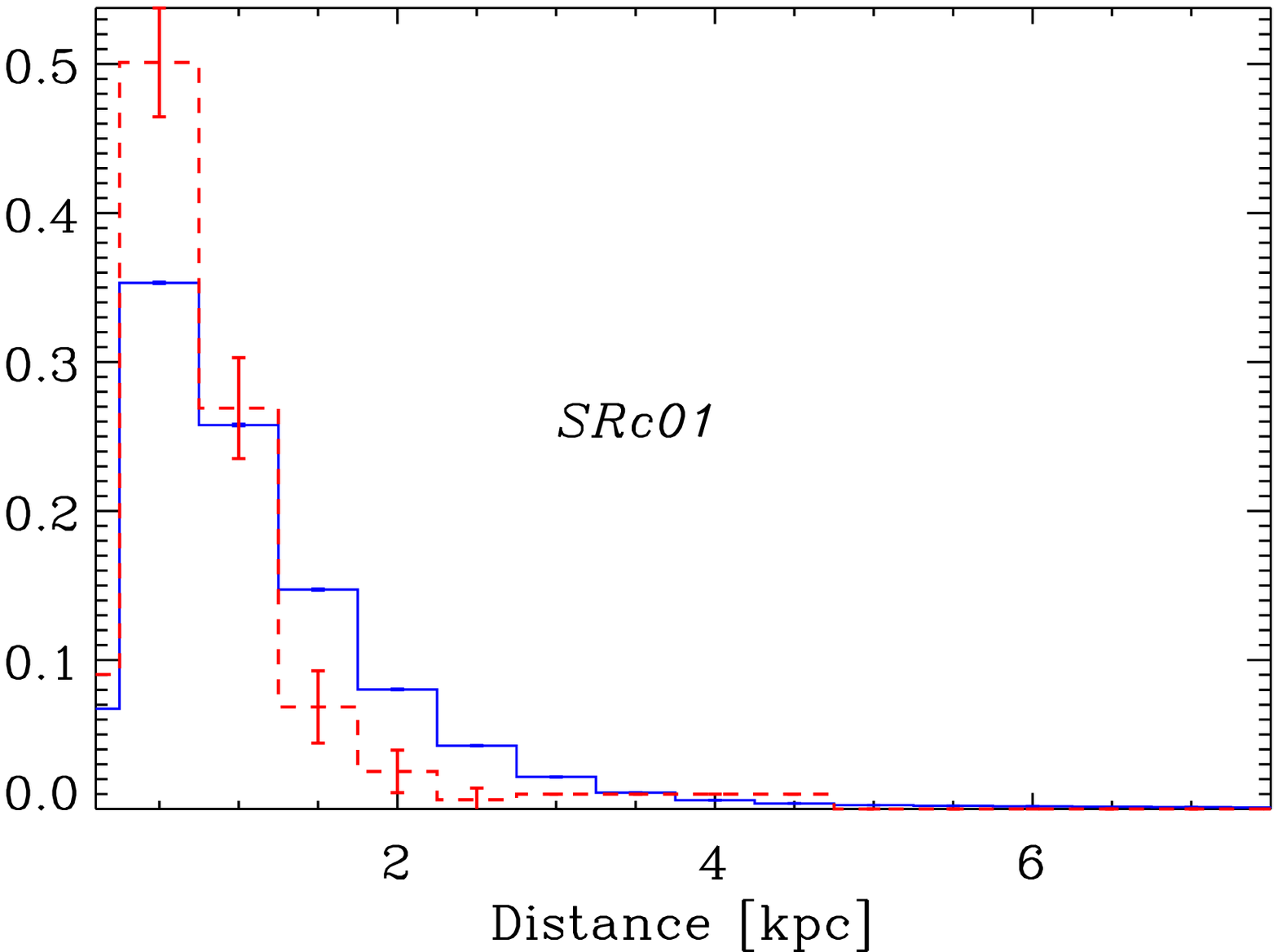}
  \caption{Distribution of the stellar distances for our observed sample (red dashed line) and the BGM1 simulation (blue). The error bars correspond to 1$\sigma$ of the Gaussian function fitting the distribution of the corresponding bin.              \label{fig:distH}}
\end{figure*}
\begin{figure*}[!t]
\centering
\includegraphics[width=0.33\textwidth]{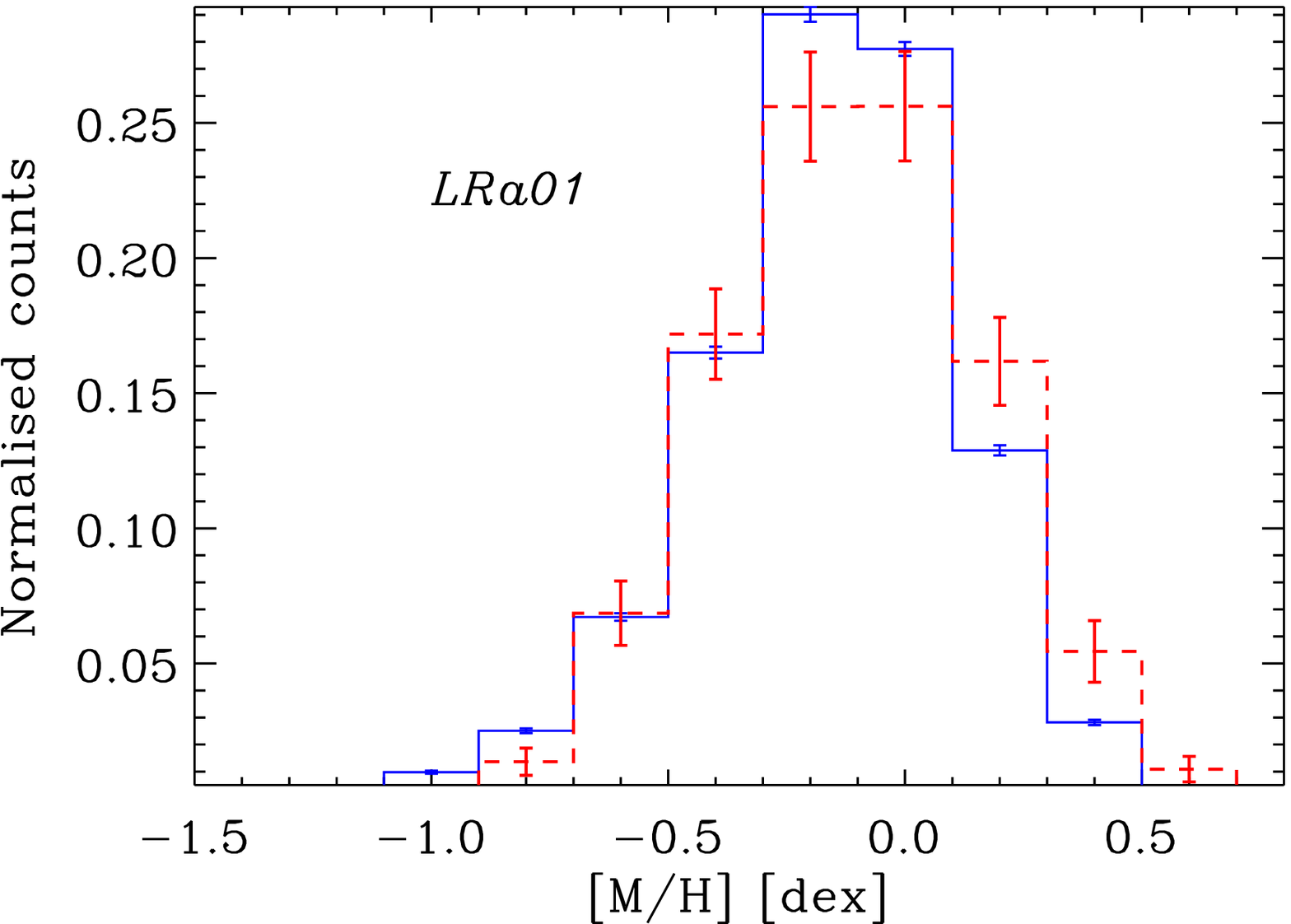}
\includegraphics[width=0.33\textwidth]{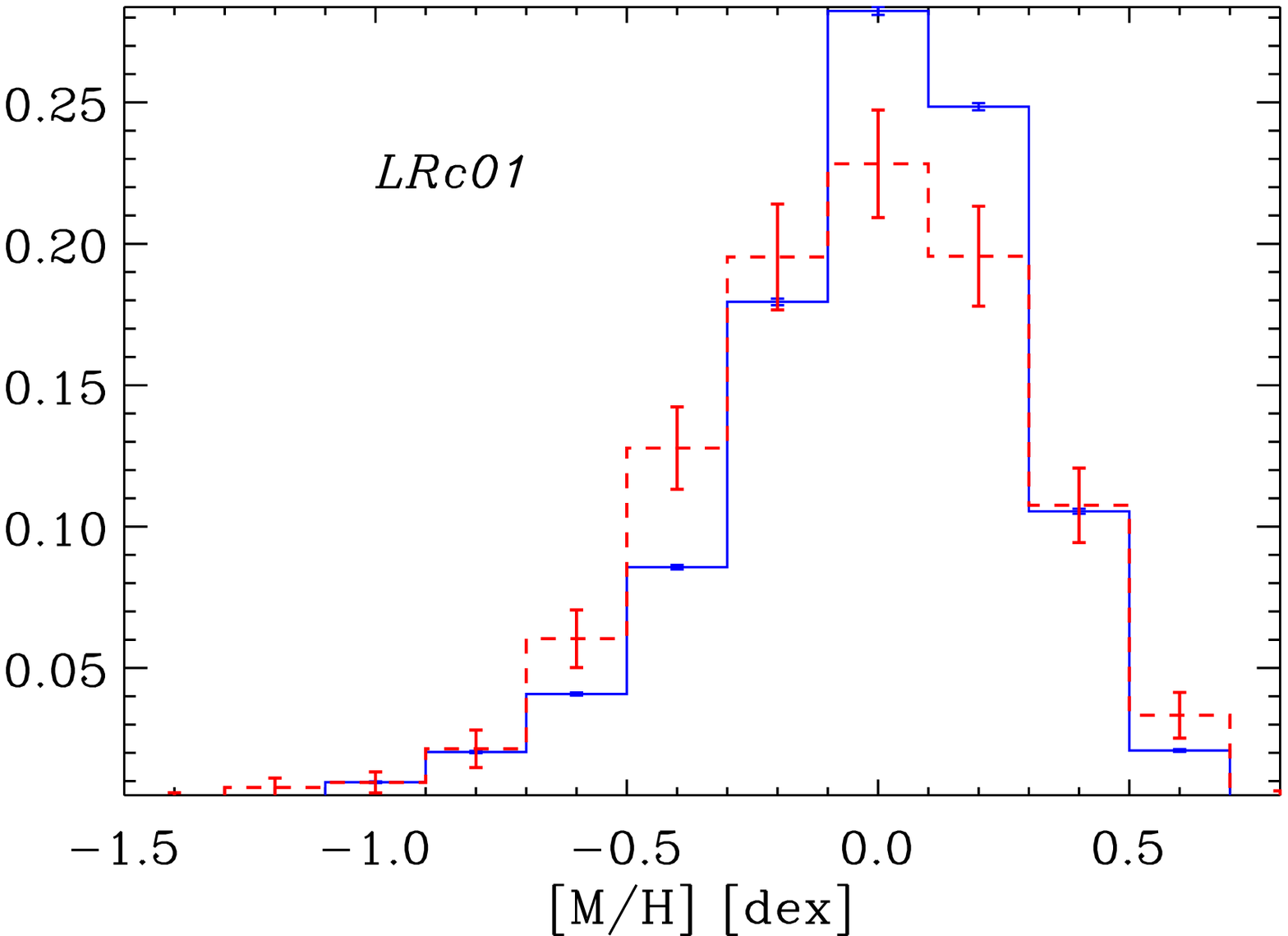}
\includegraphics[width=0.33\textwidth]{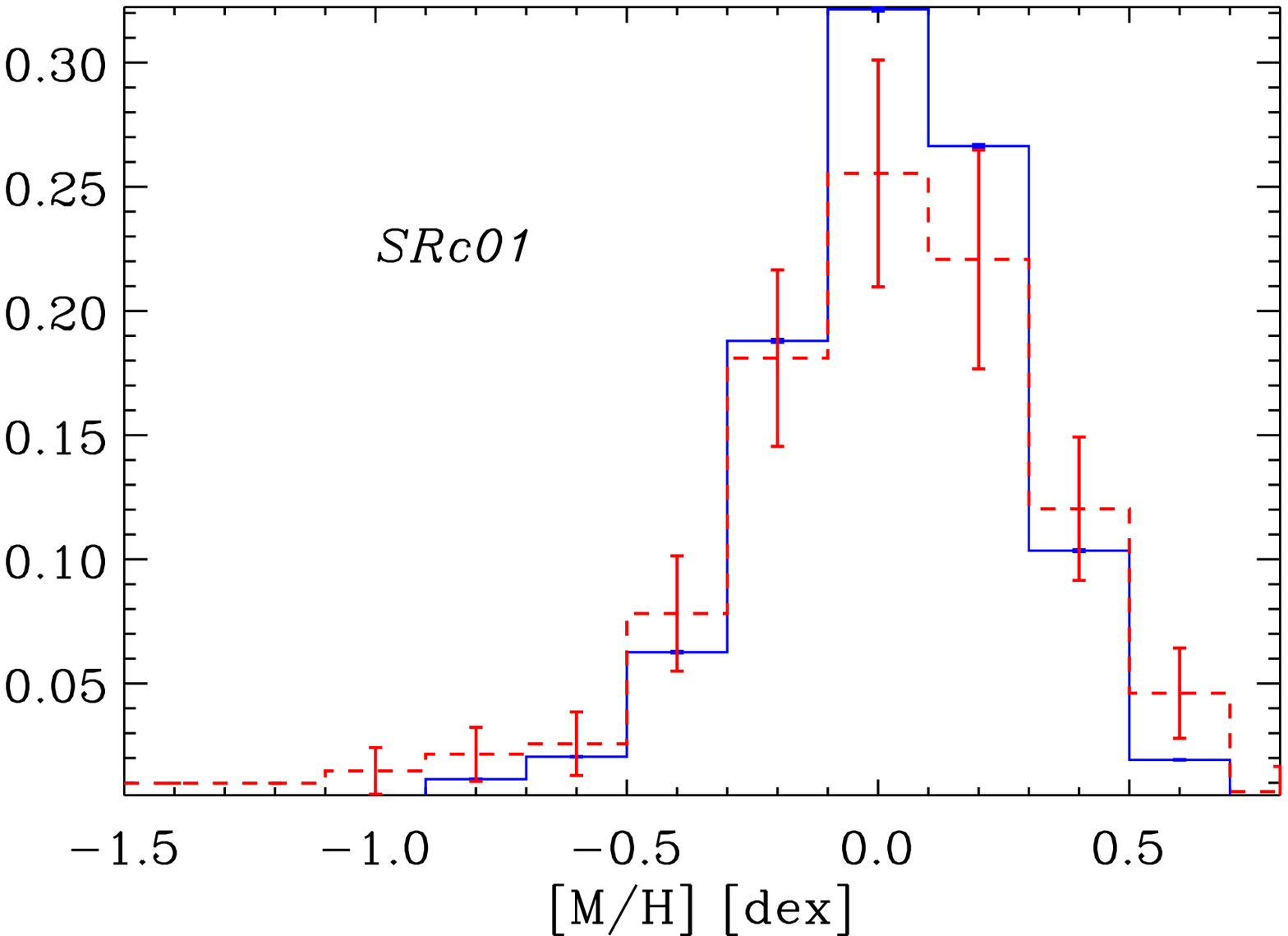}
  \caption{Distribution of the metallicity for our observed sample (red dashed line) and the BGM1 simulation (blue). The error bars correspond to 1$\sigma$ of the Gaussian function fitting the distribution of the corresponding bin. }
             \label{fig:metaHB}
\end{figure*}

\subsection{Velocity components}
\label{sect:kinematics}

To build statistically relevant distributions for the velocity components, we applied the same procedure as described above for the stellar distance distributions (2\,000 Monte-Carlo realisations). 
We propagated the uncertainty on the distance described in section \ref{sec:dist} and we also assumed a typical uncertainty of 4~mas year$^{-1}$, and 0.3~\kms, for the proper motion and radial velocity, respectively, in agreement with the values found for our sample {(see section \ref{sec:kine})}.
We also performed other requests to the Besan\c con Galatic model towards the different studied directions
taking the uncertainties we have in our data into account . Then we applied the same procedure to build the distribution of kinematical parameters for BGM1 simulated stars without adjusting any bias of the BGM simulations. 

\begin{table*}[!h]
  \caption[]{Kinematical results for the observations and the BGM1 biased simulations. The unit is \kms. \label{tab:kine}}
  \centering
{\begin{tabular}{lrrrrrr}
           \hline
           \hline
           \noalign{\smallskip}
            &{$<U>$} &{$\sigma_{U}$} &{$<V>$} &{$\sigma_{V}$} &{$<W>$} &{$\sigma_{W}$} \\
           \noalign{\smallskip}
           \hline
           \noalign{\smallskip}
\emph{LRa01}&$-$33.6$\pm$1.5&40.3$\pm$1.2&$-$20.0$\pm$1.5&34.7$\pm$1.3&1.3$\pm$1.6&35.1$\pm$1.4\\
\emph{LRa01}$_{\rm BGM1}$&$-$27.8$\pm$0.2&35.0$\pm$0.1&$-$16.6$\pm$0.2&33.8$\pm$0.2&$-$5.5$\pm$0.2&31.4$\pm$0.2\\
\emph{LRc01}&16.7$\pm$2.1&50.9$\pm$1.7&$-$17.5$\pm$2.0&45.2$\pm$1.6&$-$13.5$\pm$2.1&45.8$\pm$1.8\\
\emph{LRc01}$_{\rm BGM1}$&21.1$\pm$0.1&48.4$\pm$0.1&$-$17.3$\pm$0.1&38.8$\pm$0.1&$-$5.3$\pm$0.1&36.9$\pm$0.1\\
\emph{SRc01}&11.9$\pm$2.6&42.2$\pm$1.6&$-$6.8$\pm$1.9&25.6$\pm$1.5&$-$7.8$\pm$1.9&22.4$\pm$1.5\\
\emph{SRc01}$_{\rm BGM1}$&10.1$\pm$0.1&40.1$\pm$0.0&$-$12.9$\pm$0.0&27.5$\pm$0.0&$-$7.0$\pm$0.0&24.5$\pm$0.0\\
\hline\end{tabular}
}
\end{table*}
Figure~\ref{fig:kine} illustrates the comparison of the velocity components for our observations with the BGM1 simulated data. For each distribution in Fig.~\ref{fig:kine}, we adjusted a Gaussian function and reported the mean value and standard deviation in Table~\ref{tab:kine}. The shape of BGM1 and the observed velocity components distributions agree within 3$\sigma$. Few cases of discrepancy larger than 3$\sigma$ can, however, be noted. 
One source of these discrepancies is the distance scale that does not agree within 3$\sigma$ and which has a direct impact on the $W$ velocity component. This effect is indeed slightly alleviated in the \emph{SRc01} direction where the agreement on the stellar distances is better for the closest stars, which represent the majority of this sample. The other noticeable difference is for the $V$ component in the \emph{SRc01} direction. It could be due to little statistics numbers in this direction. Velocity dispersions are in good agreement for the three directions.

As a conclusion, the agreement (within 3$\sigma$) between our derived kinematic parameters and those simulated with the BGM1 validates our approach and the results we achieved on the observed sample. The small differences we pointed out might be due to the presence of stellar populations that could differ from those assumed when building the observed and simulated (BGM1) samples but also to the stellar extinction laws we used because they were rather poorly constrained. 
\begin{figure*}[!h]
\centering
\includegraphics[width =0.85\textwidth]{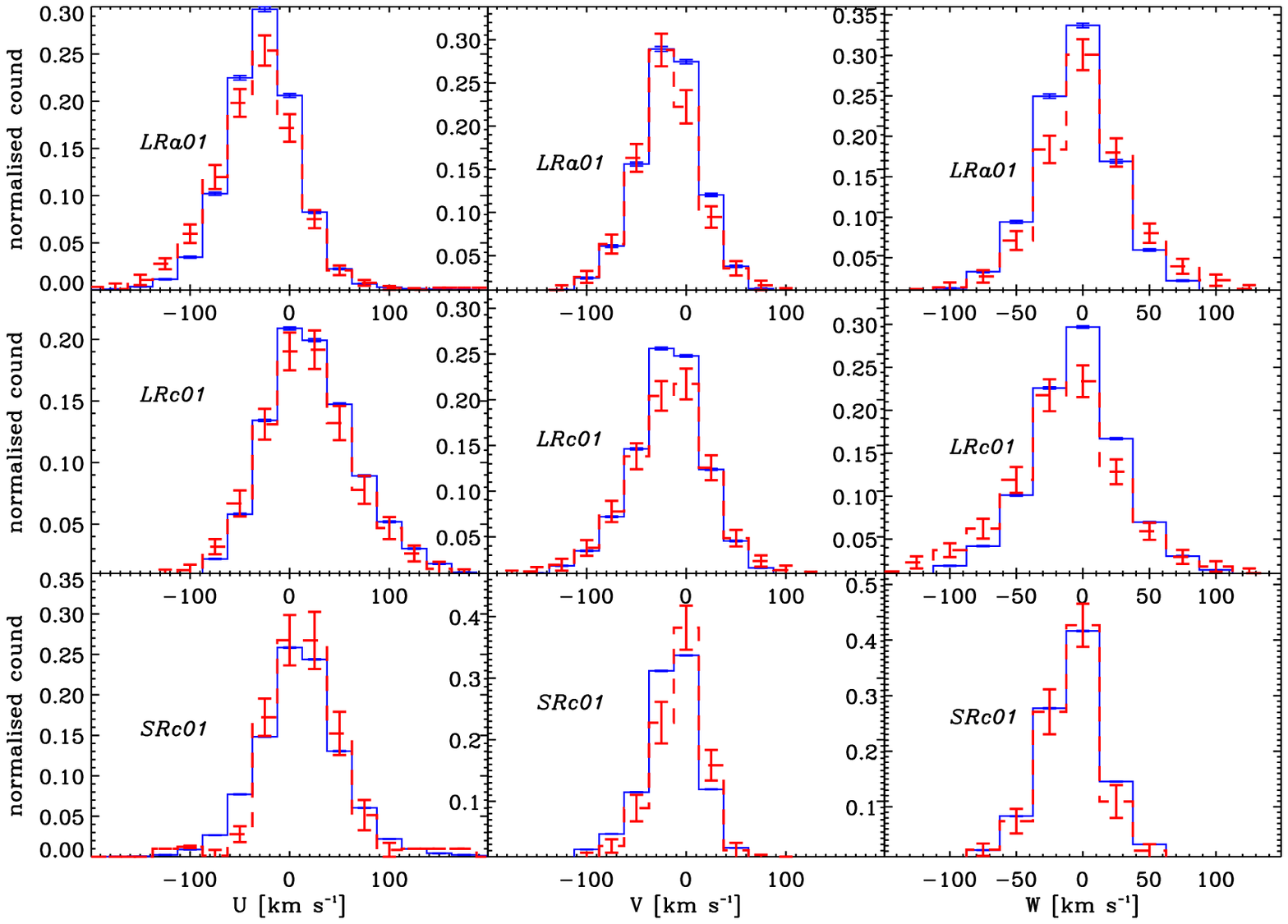}
  \caption{Kinematic comparison (three velocity components) of our observed sample with the BGM1 simulations for the three \corot~fields. The observations are represented with a red dashed line and the BGM1 data with a solid blue line.}

             \label{fig:kine}
\end{figure*}

\section{Identifying the stellar populations in the \corot~fields}
\label{Sec:stelPop}
Depending on the observed Galactic direction, any line of sight might contain a mixing of the different stellar populations (thin disc, thick disc, halo, and bulge). 
To properly interpret the observed data, we need to identify and analyse  {these different stellar  populations} separately. 
Figure \ref{fig:zr}-a) shows the height to the Galactic plane of symmetry ($Z_{gc}$) as a function of the Galactocentric radius ($R_{GC} = \sqrt{X_{GC}^{2}+Y_{GC}^{2}}$). 
Since our observations are very close to the Galactic plane, we expect our sample to be mainly composed of thin disc stars. 
However, the \emph{LRc01} direction is very likely to be contaminated by other Galactic populations, since it goes down to 6~kpc {below} the Galactic plane (see Fig.~\ref{fig:zr}-a).
We also hypothesised in the previous section that the differences between the BGM1 and the observed velocity components distributions could be due to a different mix of stellar populations.
To confirm this, we took advantage of the BGM1 simulations described in the previous section. We used the age flag given by the BGM1 to differentiate thin disc, thick disc, and halo stars (see Tab.~6). The results are summarised in Table~\ref{tab:kineBGM} (the first three lines with the age flag). According to the BGM1, {two of the} \corot~fields are mainly composed of thin-disc stars ($\sim$95\%) but the \emph{LRc01} direction indeed contains a non negligible amount (17\%) of thick-disc stars. None of the fields should contain halo stars following BGM1.
\begin{table}[!h]
\caption[]{ Age criteria associated with the different stellar populations in the BGM model \label{tab:agesBGM}}
  \centering
{      
        \begin{tabular}{cc}
           \hline
           \hline
           \noalign{\smallskip}

{Thin disc} &  0 to 10 Gyr \\
{Thick disc} & $> 10$ Gyr\\
{Halo} &\\
           \noalign{\smallskip}
           \hline

        \end{tabular}    }

\end{table}

\begin{figure*}[!h]
\centering
\includegraphics[width=0.45\textwidth]{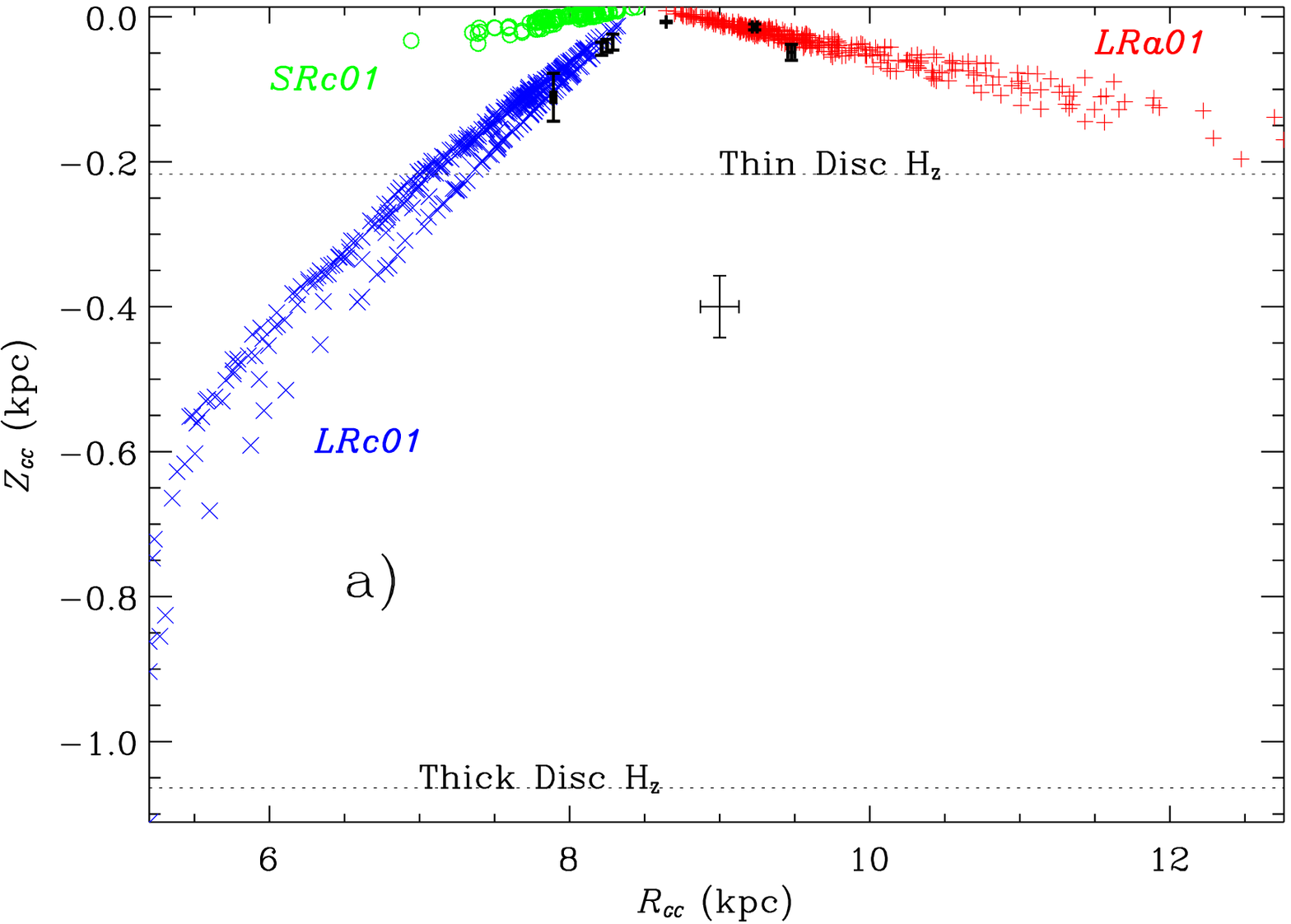}
\includegraphics[width=0.45\textwidth]{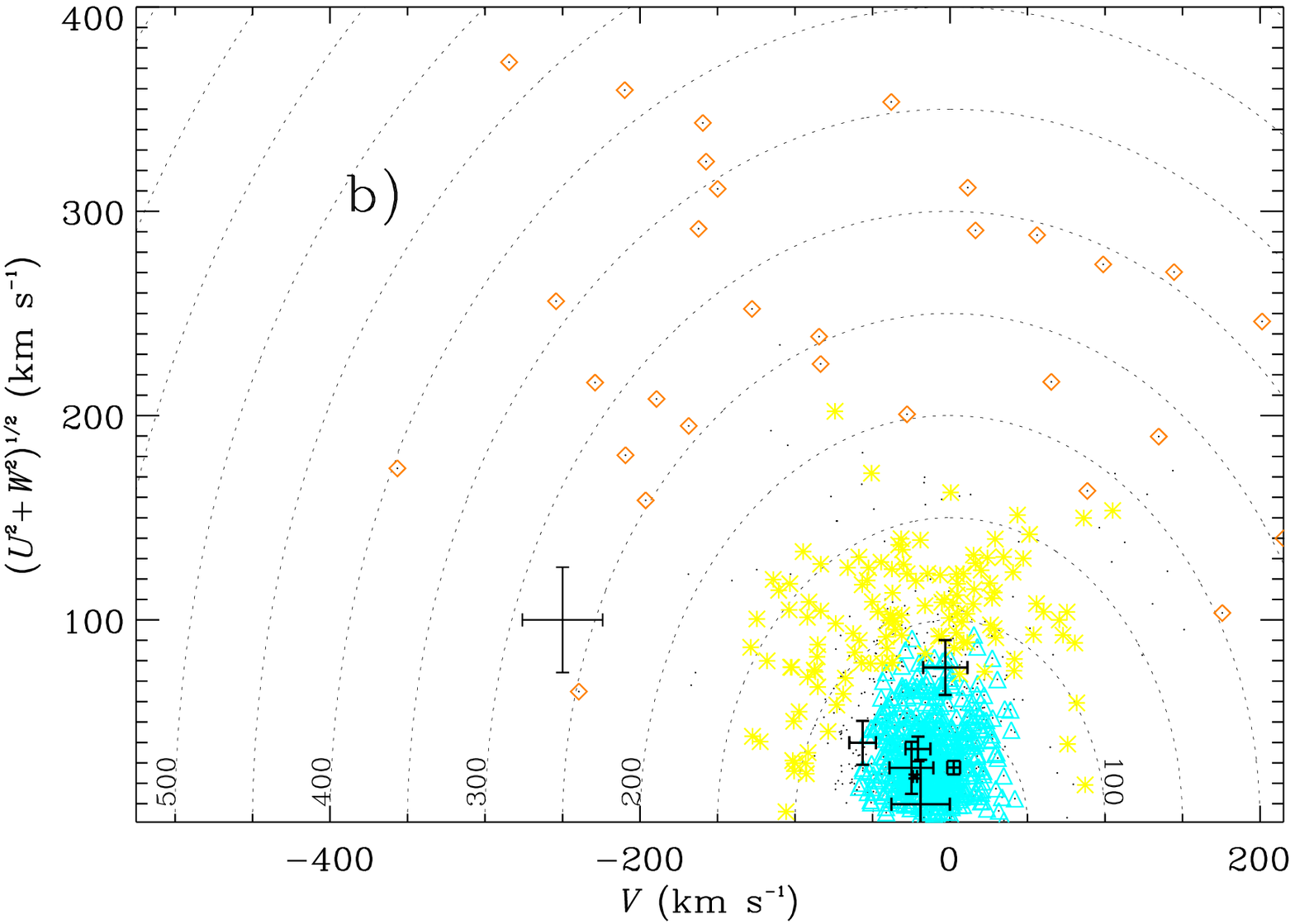}
\includegraphics[width=0.45\textwidth]{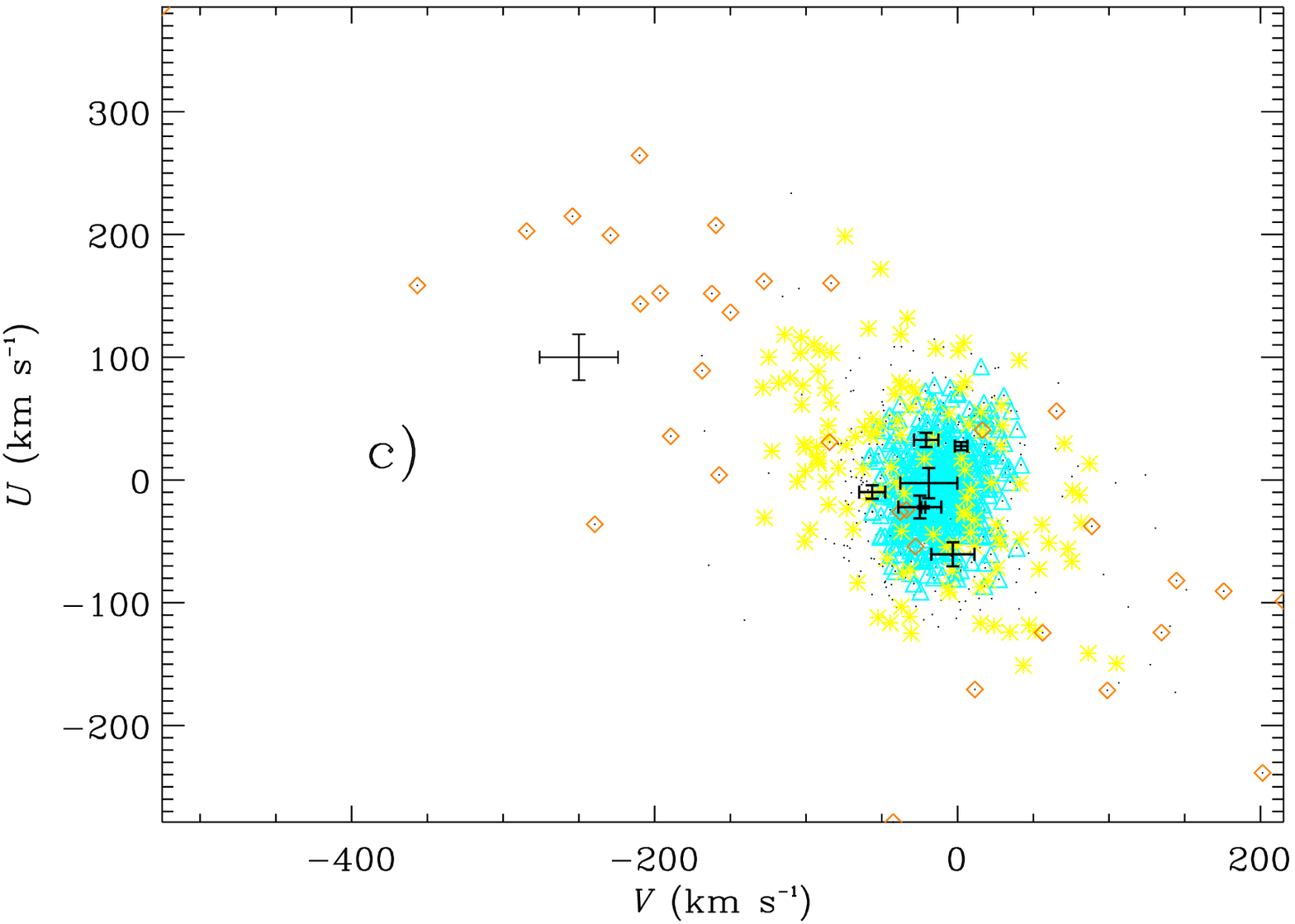}
\includegraphics[width=0.45\textwidth]{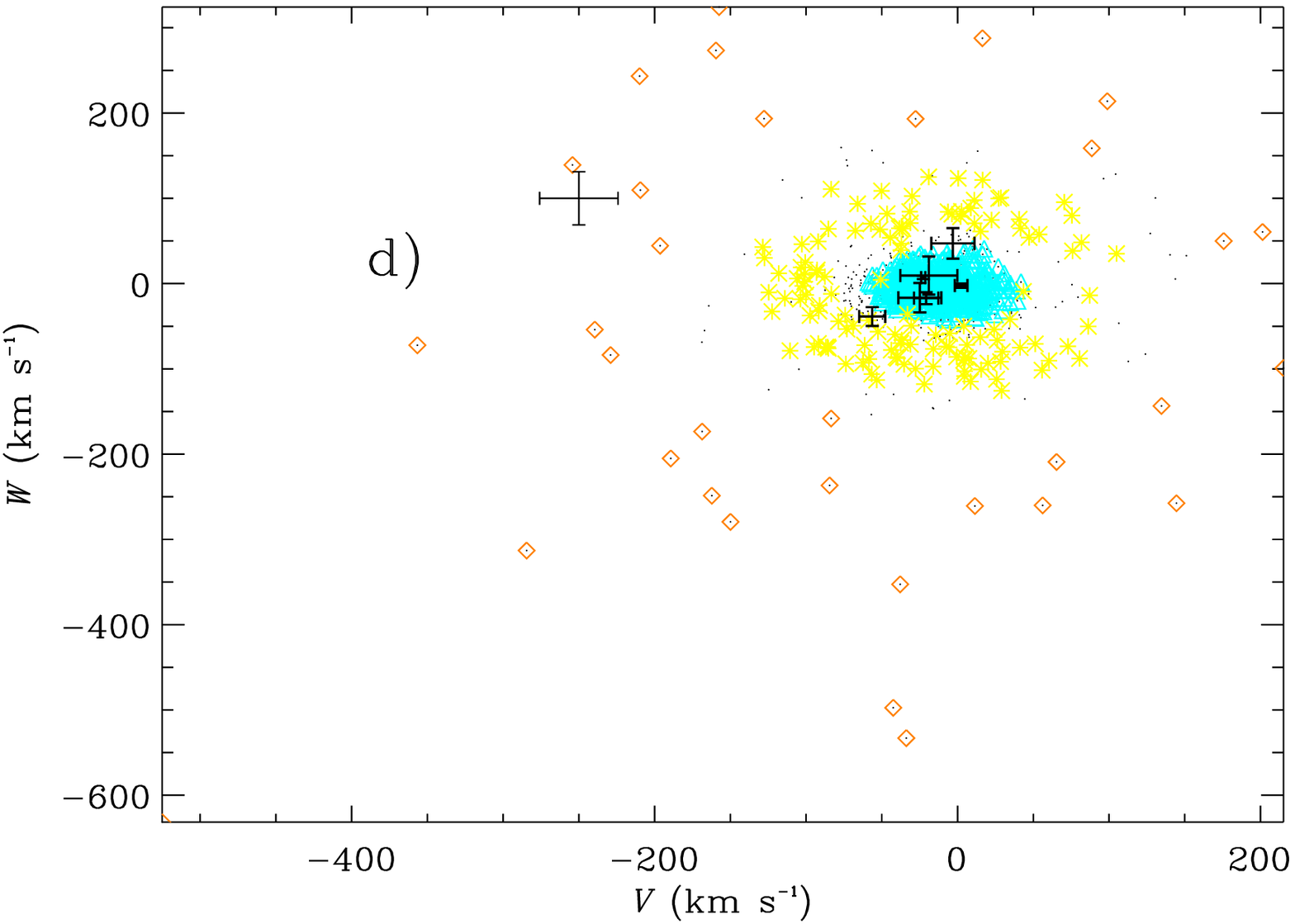}
\includegraphics[width=0.45\textwidth]{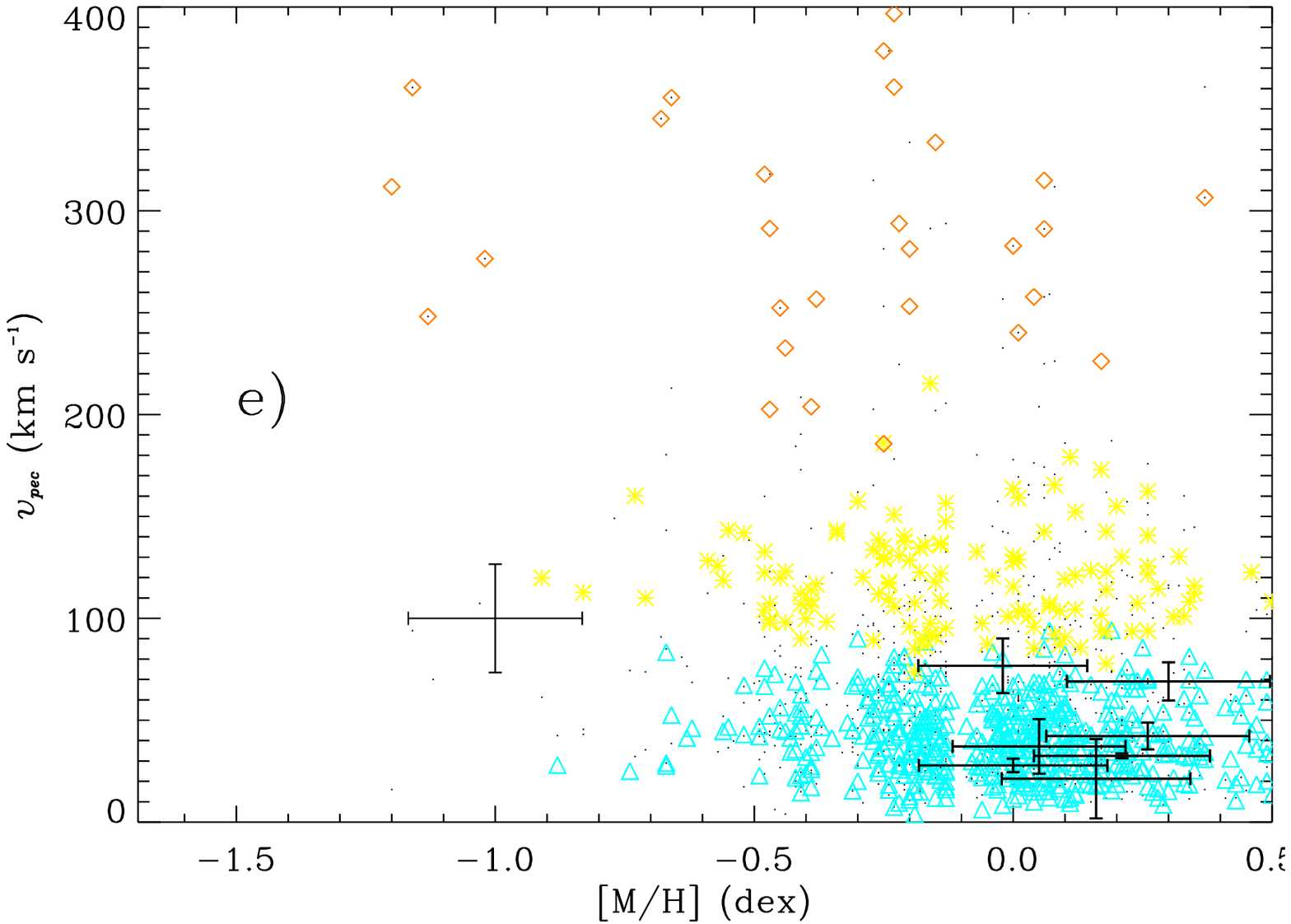}
\includegraphics[width=0.45\textwidth]{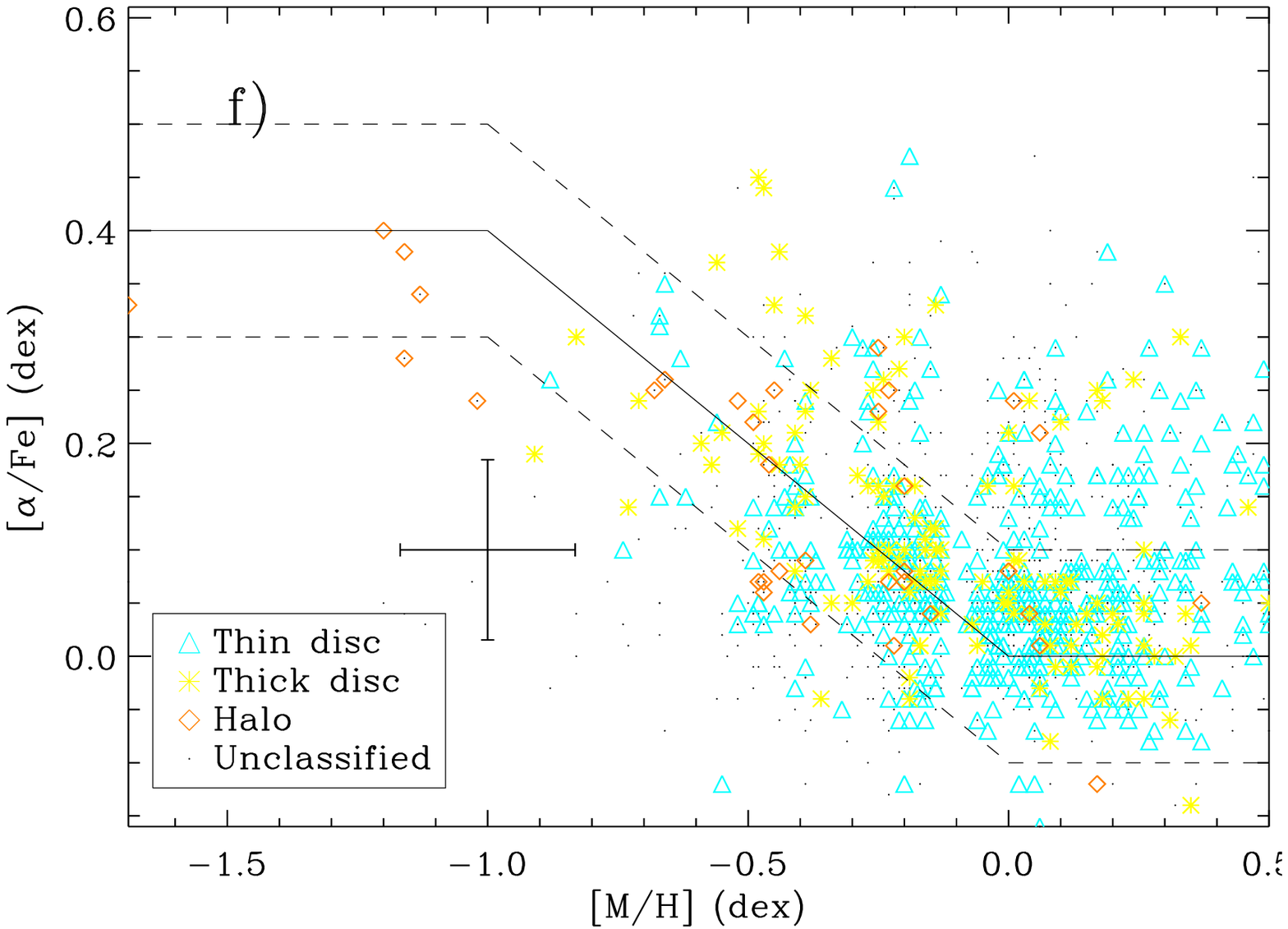}

  \caption{a) Height to the Galactic plane of symmetry ($Z_{GC}$) as a function of the Galactic radius ($R_{GC}$) for the whole sample. Blue crosses represent stars from the \emph{LRc01} direction, green circles stars from the \emph{SRc01} direction, and red plus stars from the \emph{LRa01} direction. Planet-hosting stars are represented with black thick uncertainty bars. The dotted lines correspond to the typical scale height of the thin and thick discs reported by \citet{2008A&A...480..753V}. b) Toomre diagram representing the different Galactic populations in our sample: thin disc (cyan $\triangle$), thick disc (yellow \textasteriskcentered) and halo stars, (orange $\diamond$). Dotted lines represent circles with radius every 50~\kms. c) Bottlinger diagram: $U$ versus $V$ velocities. The symbols are the same as in the Toomre diagram. The thin disc stars are focussed at $U=0$~\kms~and $V=0$~\kms. d) Kinematical heat diagram of $W$ velocities. The symbols are the same as in the Toomre diagram. The separation between the stellar populations is clear with the kinematically hottest stars being the halo stars and the coolest ones corresponding to the thin disc. e) Peculiar velocity ($v_{pec}=\sqrt{U^{2}+V^{2}+W^{2}}$) as a function of the overall metallicity. The symbols are the same as in (f). We represented in thick black the planet-hosting stars for (a) to (e). f) $\alpha-$enhancement as a function of the metallicity. The solid line corresponds to the standard law we used in our grid, the dashed line corresponds to 1$\sigma$ deviation from this law.}
             \label{fig:toomre}             \label{fig:zr}

\end{figure*}
 \begin{table}[!t]
\caption[]{Galactic population repartition in the BGM1 biased simulations. \label{tab:kineBGM}}
  \centering
{      
        \begin{tabular}{crrr}
           \hline
           \hline
           \noalign{\smallskip}
{} &
{Thin disc} &
{Thick disc} &
{Halo} \\
           \noalign{\smallskip}
           \hline
           \noalign{\smallskip}
	\emph{LRa01} BGM1 age& 96\% & 4\% & 0\% \\
    \emph{LRc01} BGM1 age& 82\% & 17\% & 1\% \\
   \emph{SRc01} BGM1 age & 95\% & 5\% & 0\% \\
            \noalign{\smallskip}
           \hline
           \noalign{\smallskip}
          
  \emph{LRa01} BGM1 kine. & 87.7\% & 2.6\% & 0.0\% \\
  \emph{LRc01} BGM1 kine. & 76.0\% & 6.2\% & 0.5\% \\
  \emph{SRc01} BGM1 kine. & 86.1\% & 3.6\% & 0.0\% \\
           \noalign{\smallskip}
           \hline
           \noalign{\smallskip}
 {   \emph{LRa01} OBS. kine. } & 58\% & 18\% & 0.3\% \\
 {  \emph{LRc01} OBS. kine.  } & 48\% & 20\% & 5\% \\
 {  \emph{SRc01} OBS. kine.  } & 82\% & 2.4\% & 0.0\% \\
           \noalign{\smallskip}
           \hline

        \end{tabular}    }
        \tablefoot{``age'' or ``kine'' correspond to the method used to separate the stellar populations: ``age'' is related to the age flag returned by the BGM1 and ``kine'' denotes classifications based on the \citet{2005A&A...433..185B} method using velocity components. For the kinematical results,  the sum of the three component does not reach 100\% because it is a  {probabilistic} way to disentangle the stellar populations, and some stars have intermediate kinematical parameters.
}
\end{table}

One independent way to  identify the stellar populations is to apply the procedure described by \citet{2005A&A...433..185B} to BGM1 simulations. This procedure combines the three Galactic velocity components to build a probability for a given star to belong to a given population. We checked that the Galactic rotation {remains} close to the solar one for the mean distances covered by our sample. We used only the three basic Galactic components (thin disc, thick disc, and halo) and we assessed those belonging to one population if the probability (defined in Eq.~\ref{eq:probabilities}) is ten times higher than the two others, which is a very strict criterion. The results shown in Table~\ref{tab:kineBGM} (three middle lines), are consistent with the age flag given by BGM, although fewer thick disc stars in the \emph{LRc01} direction are identified.

The adopted kinematic criteria are \citep{2005A&A...433..185B}:
\begin{equation}
        P = X \cdot k \cdot \exp\left(
        -\frac{U^{2}_{\rm LSR}}{2\,\sigma_{\rm U}^{2}}
        -\frac{(V_{\rm LSR} - V_{\rm asym})^{2}}{2\,\sigma_{\rm V}^{2}}
        -\frac{W^{2}_{\rm LSR}}{2\,\sigma_{\rm W}^{2}}\right),
\label{eq:probabilities}
\end{equation}
where $P$ is either $thinD$, $thickD$, or $H$, and
\begin{equation}
        k = \frac{1}{(2\pi)^{3/2}\,\sigma_{\rm U}
                                 \,\sigma_{\rm V}
                                 \,\sigma_{\rm W}}
\end{equation}
is the normalisation factor; $\sigma_{\rm U}$, $\sigma_{\rm V}$,
$\sigma_{\rm W}$ are the characteristics dispersion velocities; $V_{\rm asym}$
is the asymmetric drift; and $X$ is the fraction of observed stars in the solar neighbourhood belonging to each population.\\

\begin{table}[!b]
\centering
\caption{Characteristic velocity dispersion
        ($\sigma_{\rm U}$, $\sigma_{\rm V}$, and $\sigma_{\rm W}$) of the thin, thick disc, and halo for the equation~(\ref{eq:probabilities}).
        $V_{\rm asym}$ is the asymmetric drift and $X$ the fraction we chose.
        }
\begin{tabular}{lcccrc}
\hline \hline\noalign{\smallskip}
        & $\sigma_{\rm U}$
        & $\sigma_{\rm V}$
        & $\sigma_{\rm W}$
        & $V_{\rm asym}$ & $X$ \\
\noalign{\smallskip}
        & \multicolumn{4}{c}{(km\,s$^{-1}$)} & \%     \\
\noalign{\smallskip}
\hline\noalign{\smallskip}
   thin disc ($thinD$)   & $~~35$  & 20    & 16    & $-15$  & 94  \\
   thick disc ($thickD$) & $~~67$  & 38    & 35    & $-46$   & 6 \\
   halo ($H$)        & $160$   & 90    & 90    & $-220$   & 0.15\\
\hline
\end{tabular}
\label{tab:dispersions}
\end{table}

From the BGM age information, we search for misclassified stars using kinematical criteria. Indeed our identification of the stellar populations depends on kinematics criteria alone, and not on chemical information. Besides this, the procedure is based on solar neighbourhood data, so by applying it to our stars, we assume that these kinematical properties are also valid far from solar neighbourhood \citep{2011ApJ...735L..46B}. \citet{2010IAUS..265..300B} show that these kinematical criteria could introduce significant mixing in the identified stellar populations, but it has the advantage of not imposing an \textit{a priori} on their chemical properties.  

Applying these kinematics criteria to the BGM1 simulations, we found that thin disc stars are not contaminated by any of the two other populations. However, about 20\% of the simulated stars classified as thick disc members in the \emph{LRa01} direction are actually thin disc stars. This contamination of the thick disc population by thin disc stars is negligible ($\sim$1\%) in the \emph{SRc01} direction and goes up to $\sim$50\% in the \emph{LRc01} direction. Finally, the halo population actually comprises $\sim$10\% of thick disc stars and $\sim$20\% of thin disc stars in the \emph{LRc01} direction. 
{ 
The halo is very poorly represented in all of the simulated samples, showing that this population can be ignored altogether  in the interpretation of our observed samples.
For that reason, we decided in what follows to only characterise  the chemo-dynamical properties of  the thin disc, since obtaining a clean sample of the other components is questionable and could lead to misleading results. In addition, we stress that, whereas the BGM assumes a variation in the kinematics with the galactocentric distance, no particular adjustment of the velocity ellipsoid is needed for the spanned distances to retrieve the properties of each simulated population. Nevertheless, imposing the same kinematics as in the solar neighbourhood for the thick disc and the halo at such distances might not be realistic, so would at least bias our results kinematically. This is thus another reason for not analysing these two galactic structures.

Finally, to measure the properties of the thin disc, we used a simulation of 500~Monte-Carlo realisations, for which the observed stellar velocities were drawn from a normal distribution around their measured value  and with a standard deviation equal to the assigned uncertainty on the measurement. For each of the realisations, the kinematical criteria have been applied and a membership for each star has been associated, as described previously (Fig.~\ref{fig:toomre}-b to e,  show the results for one particular realisation: the one with the exact measured values).
The mean kinematics, gradients and metallicities were then estimated for each realisation, by fitting a Gaussian to the derived histograms. The values presented in   Table~\ref{tab:kineBGM} (bottom three lines) represent the mean proportion of each population after the Monte-Carlo simulation. 
As expected, 
}
the three directions are mainly composed of thin disc stars. The \emph{SRc01} direction is the most consistent one with the BGM1 because it is almost entirely composed of thin disc stars. As a result, the differences observed in the stellar distance and the velocity component distributions for the \emph{SRc01} are probably due to the assumed extinction laws, and the small statistics numbers. The \emph{LRa01} direction contains more thick disc stars than predicted by the BGM1, which could be one of the sources of the differences between the BGM1 and our observations.  
These two directions (\emph{SRc01} and \emph{LRa01}) are 
therefore both mainly composed of thin disc stars with a very negligible amount of thick disc and no halo stars.
Finally, the \emph{LRc01} field, however, contains a weaker thin disc contribution. It is the most mixed one and may be slightly richer in terms of thick disc stars than expected by the BGM1 simulations, although we cannot exclude that this excess thick disc population is made up of misclassified thin disc stars. Indeed, as pointed out by the analysis of the BGM1 simulated stars, although little error on the classification is expected for stars attributed to the thin disc population, the misclassification is expected to be higher for the thick disc stars, with a significant fraction of them being  thin disc stars. It is also probable that there are very few 
thick disc stars misclassified as thin disc ones since our targeted galactic directions 
contain fewer thick disc members.

All these remarks show 
that it would be safer to disentangle these stellar populations before any study devoted to the thin disc in these fields.
These results are expected from the respective mean $Z_{gc}$ and the distance range of these fields (see Fig~\ref{fig:zr}-a) and could explain the small differences between the space velocity components illustrated in Fig.~\ref{fig:distH}, \ref{fig:kine} and Table~\ref{tab:kine}.

\section{Properties of the thin disc}
\label{sec:thindisc}
According to the strict kinematics criteria and the Monte-Carlo simulations described in the previous section (bottom of Table~\ref{tab:kineBGM}), there are 
 on average 240 thin disc stars out of 404 stars observed in the \emph{LRa01} direction ($\sim$59\%), 140 out of 286 ($\sim$49\%) in the \emph{LRc01} direction, and 52 out of 64 ($\sim$81\%) in the \emph{SRc01} direction. 
 In this section, we discuss the kinematical and chemical properties of these stars identified as thin disc population. }
When analysing the kinematics of the separated stellar populations in the three fields, one has to keep in mind that we used this information to select the stars belonging to the thin disc. We recall that, based on the above discussion, 
it is probable that this thin disc sample is relatively pure, although 
biased at some level.

\subsection{Kinematics}

By selecting the thin disc stars, we ended up with the closest stars. The most distant thin disc stars are found in the \emph{LRc01} at $Z_{GC}\simeq-350$~pc, which is compatible with what is generally admitted {for} the scale of height {of the thin disc} \citep[see Fig.~\ref{fig:toomre}-a and][]{2008A&A...480..753V}.
 We used the same method as described in Sect.~\ref{Sec:stelPop} to build the distributions
   {  for each Monte-Carlo realisation and to estimate the means and standard deviations of the velocity components in each direction.  
 The dispersion{s} reported in Table~\ref{tab:kinethin} are corrected from the observational errors as in \citet{Jones88}, for each realisation. As expected, the values in  Table~\ref{tab:kinethin} are 
in generally good agreement with the criteria we used for disentangling the stellar populations. In addition, the velocity dispersions appear to correspond to the oldest part of the thin disc simulated in the BGM, though for the V-component we probably underestimate the dispersion due to an overestimation of the individual error on the V measurements. 
Finally, we analysed the evolution of the mean V-velocity for the entire sample, as a function of the galactocentric radius and found a gradient of $-5.3 \pm 1.3$~km~s$^{-1}$~kpc$^{-1}$. }

\begin{table}[!t]
{ 
\caption{Kinematical and chemical characteristics of the thin disc\label{tab:kinethin}\label{tab:chemithin}\label{tab:chemi2thin}}
\centering{
\begin{tabular}{crrr}
           \noalign{\smallskip}
           \hline
           \hline
           \noalign{\smallskip}

& \emph{LRa01}& \emph{LRc01} & \emph{SRc01}\\
           \noalign{\smallskip}
           \hline
           \noalign{\smallskip}
$<N_{\star}>$ 				& $240 \pm 6$		&$140 \pm 6$		 &$52 \pm 2$\\
$<U>$              				& $-24.4 \pm 1.4$	&$4.6 \pm	2.3$		&$11.9 \pm 3.3$\\
$<V>$              				& $-18.6 \pm1.4$	&$-13.0\pm 2.3$	&$-7.4 \pm 2.6$\\
$<W>$ 					& $-0.9 \pm 1.5$	&$-6.2 \pm 2.1 $	&$-7.8 \pm 5.1$\\
$\sigma_{U}$ 				&$27.8 \pm 1.9$	&$30.3\pm 3.1$	&$29.1 \pm 4.5$\\
$\sigma_{V}$ 				&$13.0 \pm 3.2$	& $21.7 \pm 4.3$	&$13.1 \pm 4.5$\\
$\sigma_{W}$ 				&$17.2 \pm 6.8$	&$15.8 \pm 6.9$	&$12.2 \pm 4.8$\\
$<$\met$>$ 				&$-0.05 \pm 0.01$ 	&$0.07\pm 0.02$	& $0.07 \pm 0.02$\\
$\sigma_{\rm \met}$ 			& $0.23 \pm 0.01$	 &$0.28 \pm 0.02$	&$0.23 \pm 0.04$\\
$<$\alf$>$ 				&$0.06 \pm 0.01$ 	&$0.04 \pm 0.01$	 & $0.05 \pm 0.01$\\
$\sigma_{\rm[\alpha/Fe]}$ 	& $0.05 \pm 0.01$ 	&$0.07 \pm 0.01$	&$0.05 \pm 0.01$\\
           \noalign{\smallskip}
           \hline
\end{tabular}
\tablefoot{The solar motion have not been corrected here. The dispersion values, for each Monte-Carlo realisation, have all been corrected by the observational errors as in \citet{Jones88} }}
}
\end{table}

\subsection{Metallicity}

\begin{figure}[!h]
\centering{
\includegraphics[width=0.45\textwidth]{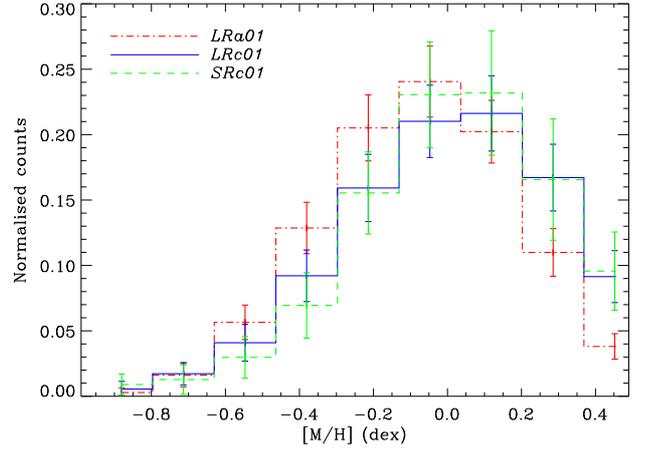}
\caption{Distribution of the overall metallicity for the thin disc sample in the three pointing directions.}
             \label{fig:histoMetaThin}}
\end{figure}

Our entire thin disc sample contains stars with metallicity ranging from $-0.88\pm0.17$ to $0.55\pm0.20$~dex. The metallicity distributions peak around 
{ 
0.07~dex for the \emph{LRc01} and \emph{SRc01} directions close to the Galactic centre and around $-0.05$~dex for the \emph{LRa01} direction, with a standard deviation of $\sim$0.25~dex} (see Fig.~\ref{fig:histoMetaThin}). 
The mean values reported in Table~\ref{tab:chemithin} agree, within 1$\sigma$, with the mean values used for the thin disc simulation in the BGM \citep{2003A&A...409..523R}  as expected according to the observational uncertainties. 
The \emph{LRa01} field exhibits a mean metallicity completely in agreement with the values for the thin disc published by \citet{2004AN....325....3F} and for the SDSS sample presented by \citet{2010IAUS..265..304A}. However, the fields in the \emph{centre} direction are slightly more metal rich.

\subsection{Radial mixing and metallicity gradient}
In Fig.~\ref{fig:histoMetaThin}, stars located towards the Galactic centre {appear to be} more metal rich than those found in the anti-centre direction. {In \citetalias{Gazzano2010}, we separated the giant from the dwarf stars using the measured \logg~and noted that this effect is clearer for giant stars, probably because the giant stars may be observed further than dwarfs. Here we use the same criterion to separate the giants (\logg$\le3.5$~dex) from dwarfs (\logg$>3.5$~dex).} These metallicity distributions can be different because of the radial and vertical metallicity gradients, radial mixing processes, and local inhomogeneities \citep{2009A&A...504...81P,2009MNRAS.396..203S,2008MNRAS.388.1175H}. We propose here to search for radial mixing signatures and to quantify the radial metallicity gradient.

\citet{2011MNRAS.412.1203N} find no correlation between $V$ velocity and metallicity, thereby constraining the degree of radial mixing \citep{2009MNRAS.396..203S,2008MNRAS.388.1175H}. We searched for such a correlation in our thin disc sample.
For each of our Monte-Carlo realisations, we have measured the correlation\footnote{Spearman's rank correlation value} between the metallicity and the V-velocity, by selecting stars in narrow ranges of galactocentric radii. Indeed, evidence of radial migration is only visible for localised samples. When larger distances are involved, correlation between metallicity and velocity appear naturally, just because the metal-richer stars in the inner galaxy rotate slower, due to the higher vertical velocity dispersion in these regions.
We considered both giants and dwarf stars in the radial ranges  $7<R_{GC}<8$~kpc, $8<R_{GC}<9$~kpc and $9<R_{GC}<10$~kpc, and find the following correlations $R=-0.028 \pm 0.077, -0.094 \pm 0.052, -0.192 \pm 0.059$, respectively. Whereas for the most inner part the correlation is not significant considering the uncertainties, this is not the case for the two other distance bins. The significance of the value at $\sim2\sigma$  is characteristic of radial migration processes in the thin disc. Indeed, as a sanity check we searched for such a correlation in the BGM1 simulation, since we recall that the BGM does not incorporate any radial migration. No such correlation was found, as expected, which is consistent with  radial-mixing processes in our observations.

\begin{figure}[!t]
\centering{
\includegraphics[width=0.49\textwidth]{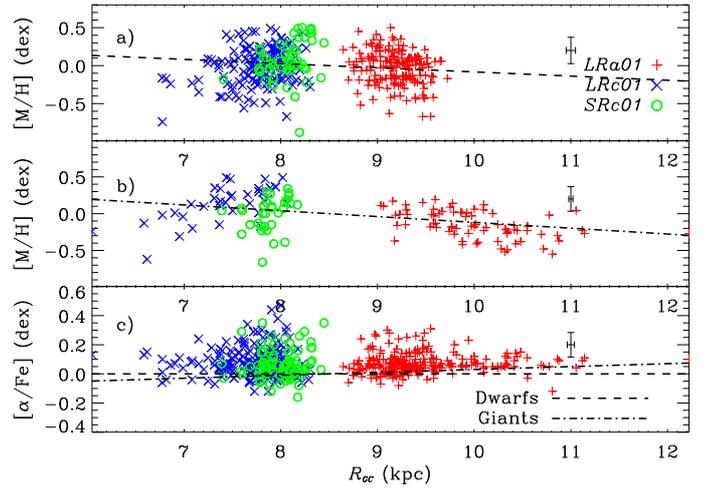}
\caption{{a) Overall metallicity as a function of Galactocentric radius for dwarf stars in the thin disc. b) Overall metallicity as a function of Galactocentric radius for giant stars in the thin disc. c) $\alpha-$enhancement as a function of Galactocentric radius. We plotted the gradient found for giant (dash-dotted line) and dwarf (dashed line) stars. Blue X represent the stars from \emph{LRc01}, green circles represent stars from \emph{SRc01}, and red + represent stars from \emph{LRa01}}.
             \label{fig:grad}}}
\end{figure}

The thin disc metallicity gradient is {also} an important input for stellar populations models to understand galactic formation and evolution \citep{2009MNRAS.396..203S}. The value of the metallicity gradient is a matter of debate, ranging from $-$0.04 to $-$0.1~dex~kpc$^{-1}$ \citep{2008MNRAS.388.1175H}. 
We explore different Galactic regions, within 1.4~kpc and 600~pc, with the giants and the dwarfs, respectively. 
{We calculated the weighted\footnote{with the external uncertainties} average of the Galactocentric radius and the metallicity for the \emph{LRa01} field, on the one hand, and the \emph{LRc01} and \emph{SRc01} fields on the other.}
We found for the dwarfs a gradient of
 $-0.053\pm0.011$~dex~kpc$^{-1}$, and for the giants, $-0.097\pm0.015$~dex~kpc$^{-1}$. 
 These results are illustrated in Figs.~\ref{fig:grad}-a) {and b) for the dwarfs and for the giants, respectively. On the one hand, the dwarf stars are known to have better determined atmospheric parameters, hence better distance and metallicity. However, the distances covered by the giants range further than for the dwarfs, which allow us to explore a larger difference in metallicity and a larger differential of distance. On the other hand, for giant stars, the uncertainties on the \logg~might result in underestimating the distance of the stars from the Sun, thus overestimating the metallicity gradient. We note, however, that when limiting it to the same distance range, the measurement of the gradient for the dwarfs and the giants marginally agree, though the same trend is apparent. For the same  radial range between 7.1~kpc and 9.3~kpc, we measure  gradients of $-0.041 \pm 0.013$~dex~kpc$^{-1}$ for the dwarfs and $-0.057 \pm 0.019$~dex~kpc$^{-1}$ for the giants. }

On the other hand, we point out that,
in order to test possible effects due to some bias in our procedure,
we measured the metallicity gradient in the BGM1 sample and showed that the result with and without uncertainties on kinematical and chemical parameters agree within error bars. 

We also tried to estimate 
the effect of our sample selection (magnitude limited sample) and individual distance uncertainties when deriving average quantities in discrete distance bins (Malmquist-type biases)
such as the gradients. For that purpose, we derived the gradients from Besan\c con simulated stars by adopting the whole sample and by randomly rejecting some faint stars (i.e. by mimicking the spectroscopic selection). It has been found that the results are only weakly affected by this effect.
Finally, we compared the metallicity gradients estimated for the BGM and BGM1 samples (direct test of the Malmquist bias).
These gradients differ by $\sim$20\% only if we consider all the stars found in BGM and BGM1.
Furthermore, the difference between the BGM and BGM1 gradients is around $\sim$35\% if we only take the
giant stars into account.
The BGM1 gradient is always the steepest one in these different tests.
This is an indication that the Malmquist bias could affect our estimated metallicity gradients, and
these gradients could be slightly overestimated. However, their values are still 
reasonably well determined, with potential biasses that do not exceed the quoted uncertainties
(see also their
comparison with other literature determinations below).

The measured metallicity gradient using 
{  dwarf stars } agree, within the error bars, with the most recent studies using Cepheid stars, with Galactocentric distances ranging from $\sim$5 to $\sim$17~kpc \citep[$-0.051\pm0.004$~dex~kpc$^{-1}$,][]{2009A&A...504...81P}, 
{ 
and slightly higher when we only consider the giants.}
Moreover, these authors show that the gradient is much steeper for the inner Galaxy ($-0.13\pm0.015$~dex~kpc$^{-1}$). We were not able to separate the inner and outer Galaxy in our sample since the scatter of each subsample is too high. 
{  
Our metallicity gradient value is also slightly higher but nevertheless compatible within $1 \sigma$} with studies based on H\textsc{ii} regions. The latest work by \citet{2006ApJS..162..346R} explores regions with $R_{GC}$ ranging from 10~kpc to 15~kpc and presents metallicity gradients\footnote{measured with oxygen, nitrogen, and sulphur abundances} ranging from $-0.046\pm 0.009$ to $-0.071\pm0.010$~dex~kpc$^{-1}$ using optical data. These authors showed that far-infrared data results in different gradient values ranging from $-0.041\pm0.009$ to $-0.085\pm0.010$~dex~kpc$^{-1}$. They finally emphasise that these results are extremely sensitive to extinction, which is one of the problems we had to face in our own study. 
Finally, observations of open clusters provide a fair estimate of metallicity and Galactocentric radius {for older stars}. Such studies find steeper gradients, {  in very } good  agreement with our estimate for the giant stars 
\citep[][]{2011MNRAS.412.1265A,2010A&A...523A..11M,2007A&A...476..217C,2002AJ....124.2693F}.
We also note that our radial metallicity gradient value for the giant stars, probing the furthest regions of the Galactic plane, may be compatible with the radial gradient used in the BGM. It is also fairly compatible with what is found using other methods. Similar studies are also devoted to other nearby galaxies such as M81, for which \citet{2010A&A...521A...3S} find an oxygen gradient of $-0.055\pm 0.02$~dex~kpc$^{-1}$, close to the value we derived for the Milky Way
{ 
dwarfs}.

\subsection{$\alpha-$enhancement}
The measurement of the \alf~ratio for different stellar populations is important for our understanding of chemical history of the Galaxy. Table~\ref{tab:chemi2thin} reports the mean and $\sigma$ values for the three fields.
The mean \alf~is identical in the three Galactic directions, and compatible with thin disc canonical value.
We used the same method as described above to measure the radial gradient for \alf.  
We found no variation in \alf~as a function of the Galactocentric radius (see Fig.~\ref{fig:grad}-b).

\section{Stars with unusually high \alf}
\label{sec:highalf}
{Among} our sample of 754 stars, we found 110 stars with \alf~values substantially higher,~\textit{i.e.} beyond the error bars ($\sigma_{[\alpha/\rm Fe]}\simeq0.1~dex$), than the standard description based
on galactic stellar properties \citepalias[see][and solid line in Fig.~\ref{fig:toomre}-f where they can be easily identified]{Gazzano2010}. These stars are found above +$1 \sigma$ and none are found below -$1 \sigma$, which makes these
detections statistically significant.
The majority of these stars are main sequence stars with mean metallicities in the range $-$0.5 to 0.5~dex. There is, however, a greater proportion of giant stars {among this group} than in the entire sample. It is also instructive that $\sim$70\% of these stars are located in the \emph{LRc01} direction. These stars show no peculiarity in terms of velocity components, and they are found at any distance towards the \textit{centre} direction, and within the closer kiloparsec towards the \emph{LRa01} direction. They mainly belong to the thin disc (56\%). 
We performed some additional tests to further verify the \matisse\ parameters and particularly the \alf~values. We first checked for these stars that the \teff~derived by \matisse~are consistent with their observed colours in the infrared (using 2MASS photometry). We then selected those among these stars with the best spectra  (\sn$\ge20$, \textit{i.e.} 28 stars). We performed a visual check for these stars, and we also searched for the synthetic spectrum in the grid that is the most similar to the observed one by minimising the $\chi^{2}$ between observed and synthetic spectra. This allowed us to confirm the \matisse~parameters for eleven of these stars. Another ten of these spectra present a slightly better fit with atmospheric parameters that re different from \matisse\ ones, but the difference never exceeds a step of the grid and are still pointing to $\alpha-$enhanced abundances. The seven remaining spectra are not adjusted with the parameters of \matisse, nor with those of the fit based on the minimum $\chi^{2}$. The examination of the cross-correlation function for these seven spectra has revealed that these stars might be spectroscopic binaries not detected by \citet{Loeillet2008FLAMES}. Therefore from these 28 rather good \sn~spectra, we can confirm the peculiar \alf~for 21 stars.
{Finally, we extended the $\chi^2$ test to all of these $\alpha$-peculiar stars: we searched for the grid point minimising the $\chi^{2}$ with the observed spectrum. We found that 90\% of these spectra have $\chi^2$ parameters $\alpha-$enhanced.}

All these checks therefore confirm the non-standard \alf~value of this subsample of stars. 
We therefore {might} have detected metal-rich stars with unexpected {peculiar} high \alf~values that contradict the current scenarios of evolution and chemistry of the galactic disc. For example, it has been proposed that
bulge stars could be $\alpha-$enhanced because they are supposed to form quickly \citep{1994ApJS...91..749M,1990ApJ...365..539M},
although more recent studies seem to contradict this point (see, for instance, \citet{Bensby2011}). Our $\alpha-$peculiar stars might find an origin in the Galactic bulge and have migrated to the solar neighbourhood \textit{via} radial mixing. These stars would deserve further studies to estimate their chemical abundances in detail and to disentangle their origin.\footnote{A few months after the submission of this article, a discovery of similar $\alpha$-enhanced metal-rich stars has been reported by
\citet{Adibekyan}, thus confirming our detections.} 

\section{Properties of the planet-hosting stars in the targeted directions}
\label{sec:planets}
\begin{table*}
\caption{Atmospheric parameters for the planet-hosting stars detected by \corot~in our targetted directions\label{tab:planets}}
\centering{
\begin{tabular}{cclrrrrrrrrc}
           \noalign{\smallskip}
           \hline
           \hline
                     \noalign{\smallskip}
                     \corot~field & 
{\corot~ID} &
  {Planet Name} &
  {\teff} (K)&
  {$\sigma_{\rm T_{eff}}$} &
  {\logg} &
  {$\sigma_{\log~g}$} &
  {\met} &
  {$\sigma_{\met}$} &
  {\alf} &
  {$\sigma_{\rm [\alpha/Fe]}$} & Reference\\
           \noalign{\smallskip}
           \hline
                     \noalign{\smallskip}
\emph{LRc01} & 101206560 & CoRoT-2b & 5625.0 & 163.5 & 4.3 & 0.335 & 0.0 & 0.182 & 0.0 & 0.217&(1)\\
\emph{LRc01} &  101368192 & CoRoT-3b & 6740.0 & 178.7 & 4.22 & 0.278 & -0.02 & 0.163 & -0.05 & 0.115 & (2)\\
\emph{IRa01} &  102912369 & CoRoT-4b & 6190.0 & 126.2 & 4.41 & 0.274 & 0.05 & 0.167 & 0.0 & 0.217 & (3)\\
\emph{LRa01} &  102708694 & CoRoT-7b & 5319.0 & 120.4 & 4.76 & 0.186 & 0.21 & 0.17 & 0.05 & 0.071 & (4)\\
\emph{LRc01} &  101086161 & CoRoT-8b & 5080.0 & 143.4 & 4.58 & 0.201 & 0.3 & 0.196 & -0.04 & 0.099 & (5)\\
\emph{LRc01} &  100725706 & CoRoT-10b & 5075.0 & 140.7 & 4.65 & 0.209 & 0.26 & 0.196 & 0.06 & 0.139 & (6)\\
\emph{LRa01} &  102671819 & CoRoT-12b & 5675.0 & 136.8 & 4.52 & 0.281 & 0.16 & 0.182 & -0.06 & 0.138& (7)\\
           \noalign{\smallskip}
           \hline
\end{tabular}}
\tablebib{(1) \citet{2008A&A...482L..21A,2008A&A...482L..25B} ; (2) \citet{2008A&A...491..889D} ; (3) \citet{2008A&A...488L..47M} ; (4) \citet{Gazzano2010} ; (5) \citet{2010A&A...520A..66B} ; (6) \citet{2010A&A...520A..65B} ; (7) \citet{2010A&A...520A..97G}
}
\end{table*}

\begin{table*}
\caption{Absorption, distance, and geocentric coordinates for the \corot~planet-hosting stars in our fields\label{tab:planets1}}
\centering{
\begin{tabular}{lrrrrrrrrrr}
           \noalign{\smallskip}
           \hline
           \hline
           \noalign{\smallskip}
  \multicolumn{1}{c}{\corot~ID} &
  \multicolumn{1}{c}{$A_{J}$} &
  \multicolumn{1}{c}{$\sigma_{A_{J}}$} &
  \multicolumn{1}{c}{$D$} &
  \multicolumn{1}{c}{$\sigma_{D}$} &
  \multicolumn{1}{c}{$X$} &
  \multicolumn{1}{c}{$\sigma_X$} &
  \multicolumn{1}{c}{$Y$} &
  \multicolumn{1}{c}{$\sigma_Y$} &
  \multicolumn{1}{c}{$Z$} &
  \multicolumn{1}{c}{$\sigma_Z$} 
\\
  &
  \multicolumn{2}{c}{} &
  \multicolumn{2}{c}{(pc)} &
  \multicolumn{6}{c}{(kpc)} \\

           \noalign{\smallskip}
           \hline
                     \noalign{\smallskip}
  CoRoT-2b & 0.1436 & 0.2047 & 275.387 & 84.718 & 0.214 & 0.066 & 0.169 & 0.052 & -0.035 & 0.011\\
  CoRoT-3b & 0.2659 & 0.1855 & 789.243 & 231.549 & 0.622 & 0.182 & 0.473 & 0.139 & -0.111 & 0.033\\
  CoRoT-4b & 0.1156 & 0.1709 & 862.254 & 224.761 & -0.721 & 0.188 & -0.472 & 0.123 & -0.014 & 0.0040\\
  CoRoT-7b & 0.0582 & 0.1875 & 171.124 & 28.299 & -0.143 & 0.024 & -0.093 & 0.015 & -0.0070 & 0.0010\\
  CoRoT-8b & 0.2793 & 0.2193 & 324.957 & 70.942 & 0.253 & 0.055 & 0.199 & 0.044 & -0.04 & 0.0090\\
  CoRoT-10b & 0.3672 & 0.2204 & 366.143 & 77.352 & 0.289 & 0.061 & 0.221 & 0.047 & -0.044 & 0.0090\\
  CoRoT-12b & 0.1585 & 0.2006 & 1144.211 & 250.813 & -0.957 & 0.21 & -0.625 & 0.137 & -0.049 & 0.011\\
           \noalign{\smallskip}
           \hline
\end{tabular}
}
\end{table*}

\begin{table*}
\caption{Velocity components for the \corot~planet-hosting stars in our directions. \label{tab:planets2}}
\centering{
\begin{tabular}{lrrrrrr}
           \noalign{\smallskip}
           \hline
           \hline
           \noalign{\smallskip}
  \multicolumn{1}{c}{\corot~ID} &
  \multicolumn{1}{c}{$U$} &
  \multicolumn{1}{c}{$\sigma_U$} &
  \multicolumn{1}{c}{$V$} &
  \multicolumn{1}{c}{$\sigma_V$} &
  \multicolumn{1}{c}{$W$} &
  \multicolumn{1}{c}{$\sigma_W$} \\ 
    &
  \multicolumn{6}{c}{(\kms)} 
\\
           \noalign{\smallskip}
           \hline
                     \noalign{\smallskip}
  CoRoT-2b & 27.6 & 3.31 & 2.356 & 4.158 & -2.38 & 2.467  \\ 
  CoRoT-3b & -60.514 & 9.77 & -3.056 & 14.263 & 47.093 & 17.883  \\ 
  CoRoT-4b & -22.014 & 9.278 & -24.873 & 14.143 & -16.592 & 17.273 \\ 
  CoRoT-7b & -22.246 & 0.845 & -22.628 & 1.366 & 7.109 & 1.778  \\ 
  CoRoT-8b & -9.805 & 5.54 & -56.366 & 8.562 & -38.62 & 11.007 \\ 
  CoRoT-10b & 32.593 & 5.787 & -20.686 & 8.052 & -17.141 & 6.993  \\ 
  CoRoT-12b & -2.56 & 12.293 & -18.959 & 18.825 & 9.388 & 22.324  \\ 
           \noalign{\smallskip}
           \hline
\end{tabular}
}
\end{table*}

In \citetalias{Gazzano2010}, we showed that the de-biased metallicity distributions combined with planet-occurrence probability laws, giant-planet period distribution, and geometric probability of transit, provided a number of detections agree with what was detected by \corot~in these fields. 
\corot~detected seven planetary systems in the Galactic directions targetted in this study. We retrieved the atmospheric parameters from the {literature} (see Table~\ref{tab:planets}), except for \corot-7b for which we had {already} derived parameters in \citetalias{Gazzano2010}. These atmospheric parameters agree, within the error bars, with the values published in \citet{2009A&A...506..287L} and \citet{2010A&A...519A..51B}. From these parameters, we used the same methodology as for our \giraffe/\flames~sample to derive the kinematical information. The distance derived here is generally in good agreement with published values for these planet-hosting stars. {We represented these planet-hosting stars in Fig.~\ref{fig:toomre}-a).}
We also applied the same kinematical criteria to identify the stellar population these stars might belong to. We find that the planet-hosting stars mainly belong to the thin disc. However, the kinematics of \corot-3b and 8b did not allow us to classify them  if we adopt the same strict criterion as for our \giraffe/\flames~sample. Indeed, the probability of belonging to the thin disc is only 2.5 times greater than the probability of belonging to the thick disc for \corot-3b and only 0.5 times greater for \corot-8b. It is particularly interesting to note that these two stars are located in the \emph{LRc01} field, which is the field that is the most mixed up in terms of populations and ages according to BGM. However, the $Z_{GC}$ for these stars is compatible with thin disc stars, as shown in Fig.~\ref{fig:toomre}-a). We also find a clear correlation (R=$-$0.71)  between the $V$ velocity component and the metallicity for these planet-hosting stars. Although the statistics is low, this tends to favour a radial mixing origin of planet metallicity correlation as proposed by \citet{2009ApJ...698L...1H}.

\section{Conclusion\label{SecDis}}

We computed stellar distances for the 754 \corot~stars with atmospheric parameters derived using the \matisse~algorithm \citep{Gazzano2010}.
This allowed us to map the Galactic kinematics and chemistry in three Galactic directions observed by \corot. All the results are available electronically through the \exodat~database \citep{2009AJ....138..649D}.

Using the kinematical criteria described by \citet{2005A&A...433..185B}, 
we identified thin disc and thick disc stars in our observed sample. The procedure was first validated on stars simulated with the Besan\c con Galactic model which allowed us to investigate the possible limitation of the classification method. We found that the proportion of thin disc stars misclassified as thick disc ones should be significant especially in the \emph{LRc01}. Our results for the observed sample show that our kinematical properties are in good agreement with predictions of the Besan\c con model, which is, however, very sensitive to the stellar extinction law assumed. The \emph{LRa01} and \emph{LRc01} directions contain fewer thin disc stars and more stars suspected of belonging to the thick disc population than predicted. This could explain the small differences observed between space-velocity component distributions.

The adopted selection criteria also allowed us to build a clean sample of thin disc stars to study the properties of this stellar population. Combining the velocity dispersions and the chemistry, 
we  found a correlation between the $V$-velocity  component and \met, for two of the three considered radial bins, which could be a clue of radial mixing \citep{2011MNRAS.412.1203N,2009MNRAS.396..203S,2008MNRAS.388.1175H}. With dwarf stars, we also measured a thin disc radial metallicity gradient of $-0.053\pm0.011$~dex~kpc$^{-1}$, which is consistent with the most recent published values \citep{2010IAUS..265..317M}.

Our analysis also shows the presence of stars in our sample with unexpected high \alf~values at rather high metallicities. This stellar population might have been formed in the Galactic bulge and migrated up to solar neigbourhood \citep{1994ApJS...91..749M,1990ApJ...365..539M}.  Although our spectroscopic data are not appropriate for carrying out
the detailed analysis that such stars would require to better understand their origin, the existence of
this puzzling population has nevertheless been recently confirmed by \citet{Adibekyan}. 

Applying the same methodology to the planet-hosting stars discovered by \corot~in the targeted directions, we showed that they mainly belong to the thin disc stellar population. This probably means that, at least in these three fields, \corot~has detected planets only around thin disc stars. We also found a correlation (R=$-$0.71)  between the $V$ velocity component and the metallicity for these planet hosting stars, suggesting a radial mixing history for these stars.

This work is one of the first studies of the stellar populations in the Galactic plane not limited to the strict solar neighbourhood, and based on good statistics although still limited. This demonstrated the potential of multi-fibre instrument like \flames/\giraffe, combined with automatic analysis tools like \matisse \, for Galactic physics analyses. It would be interesting to complete this study in other Galactic directions, observed or not by \corot, and to combine these results with the richness of the information that can be derived from \corot~light-curves analysis. For instance, one could combine the atmospheric parameters with the light curve parameters, \textit{i.e.} degree and scale of variability, to search for links between the stellar parameters and the photometric variations of the star.

\begin{acknowledgements}
Computations have been done on the ``Mesocentre SIGAMM''
machine, hosted by Observatoire de la C\^ote d'Azur.
We thank C. Gry, L. Deharveng, S. Boissier, and C. Schimd for fruitful discussions.
We sincerely thank the anonymous referee for a careful reading and constructive remarks.
\end{acknowledgements}

\bibliographystyle{aa.bst}
\bibliography{flames2}

\end{document}